\documentclass[preprint,12pt,3p]{elsarticle}

\usepackage{amssymb}
\newdefinition{rmk}{Remark}

\usepackage{hyperref}
\usepackage{graphicx}

\usepackage{epstopdf} 

\usepackage{caption}
\DeclareGraphicsExtensions{.eps,.mps,.pdf,.jpg,.png}
\graphicspath{{figures/}{../figures/}}

\usepackage{fancyhdr}

\usepackage{amsmath}
\usepackage{amsfonts}

\usepackage{dsfont}

\usepackage{array}
\newcolumntype{C}[1]{>{\centering\let\newline\\\arraybackslash\hspace{0pt}}m{#1}}

\newcommand{\dd}{{\rm d}}

\renewcommand{\vec}[1]{{\mathbf #1}}

\newcommand{\x}{{\vec{x}}}

\newcommand{\z}{{\vec{z}}}
\newcommand{\g}{{\vec{g}}}
\newcommand{\q}{{\vec{q}}}
\newcommand{\p}{{\vec{p}}}

\newcommand{\A}{{\vec{A}}}
\newcommand{\C}{{\vec{C}}}

\newcommand{\I}{{\vec{I}}}

\newcommand{\RB}{{\mathrm{\mathbf{R}}}}

\newcommand{\WB}{{\mathrm{\mathbf{W}}}}
\newcommand{\xiB}{{\boldsymbol{\xi}}}
\newcommand{\etaB}{{\boldsymbol{\eta}}}

\newcommand{\SigmaB}{{\boldsymbol{\Sigma}}}

\usepackage{mathrsfs}

\usepackage{soul}
\usepackage{color}

\usepackage[titletoc,toc,title]{appendix}

\begin{document}

\begin{frontmatter}

\title{Accurate and robust splitting methods for the generalized Langevin equation with a positive Prony series memory kernel}

\author[label1]{Manh Hong Duong}
\address[label1]{School of Mathematics, University of Birmingham, Edgbaston, Birmingham, B15 2TT, United Kingdom}

\ead{h.duong@bham.ac.uk}

\author[label1,label2]{Xiaocheng Shang\corref{cor1}}
\address[label2]{The Alan Turing Institute, British Library, 96 Euston Road, London, NW1 2DB, United Kingdom}

\ead{x.shang.1@bham.ac.uk}

\cortext[cor1]{Corresponding author.}

\begin{abstract}
  We study numerical methods for the generalized Langevin equation (GLE) with a positive Prony series memory kernel, in which case the GLE can be written in an extended variable Markovian formalism. We propose a new splitting method that is easy to implement and is able to substantially improve the accuracy and robustness of GLE simulations in a wide range of the parameters. An error analysis is performed in the case of a one-dimensional harmonic oscillator, revealing that several averages are exact for the newly proposed method. Various numerical experiments in both equilibrium and nonequilibrium simulations are also conducted to compare the method with popular alternatives in interacting multi-particle systems.
\end{abstract}


\end{frontmatter}


\pagenumbering{arabic}

\section{Introduction}
\label{sec:Introduction}

It is well known that the popular Langevin dynamics is based on the assumption that there is a clear separation between the characteristic time scale of the ``massive'' particles and that of those smaller solvent particles forming the heat bath. Within this assumption, the effect of the solvent is simply reduced to an instantaneous drag force and a delta-correlated random force, thereby significantly reducing the computational cost of an ``explicit solvent'' model. However, it has been well documented that this underlying assumption breaks down in many physically compelling scenarios, in which case a generalized version of the Langevin dynamics is more suitable---it is known as the generalized Langevin equation (GLE) derived from the Mori--Zwanzig formalism~\cite{Mori1965,Zwanzig2001}. The GLE has been widely applied in a large number of applications, including molecular simulations~\cite{Hijon2010,Lei2016a,Grogan2020}, mesoscopic modeling~\cite{Cordoba2012,Jung2018}, solids~\cite{Kantorovich2008a,Kantorovich2008,Stella2014}, nuclear quantum effects~\cite{Ceriotti2009a,Ceriotti2010},
and various anomalous diffusion phenomena~\cite{Glatt-Holtz2020,McKinley2018}.

Unlike the Langevin dynamics, a temporally nonlocal drag force and a random force with nontrivial correlations are typically associated with the GLE, which make its numerical integration highly nontrivial~\cite{Hijon2010,Baczewski2013}. To be more precise, the temporally nonlocal drag force, in the form of a convolution of the momentum with a memory kernel, requires the storage of the momentum history, whose numerical evaluation at each time step can be computationally demanding. On the other hand, it can also be computationally expensive to generate a random force with nontrivial correlations, which may require the storage of a sequence of random numbers at each time step.
Although considerable effort has been devoted to developing accurate and robust numerical methods that circumvent either or both of these challenges mentioned above, it is very difficult to claim any individual method as being optimal, especially given the fact that there are such a broad range of applications where the GLE can be applied.

To this end,
in this article we focus our attention on a general form of the memory kernel, which is a sum of exponentials and also known as a positive Prony series. It has been widely used in a large number of studies in the literature~\cite{Fricks2009,McKinley2009,Ottobre2011,Baczewski2013,Pavliotis2014,Hall2016}. Moreover, as pointed out in~\cite{Abate1999,McKinley2009,Glatt-Holtz2020}, with certain choice of the parameters, a Prony series can be seen as an approximation of a power law~\cite{Kupferman2004,McKinley2018,Glatt-Holtz2020}, another primary example of the memory kernel.

The rest of the article is organized as follows. In Section~\ref{sec:Mathematical_Formulations}, we discuss the mathematical formulations of the GLE, including not only the extended variable Markovian formalism and its white noise limit but also the Markovian approximations of the GLE with a more general memory kernel. Section~\ref{sec:Numerical_Methods} reviews three popular integration methods for GLE, followed by the derivations of a new promising scheme. Error analysis on the averages in the case of a one-dimensional harmonic oscillator will also be performed. A variety of numerical experiments are performed in Section~\ref{sec:Numerical_Experiments} to compare all the schemes described in the article. Our findings are summarized in Section~\ref{sec:Conclusions}.

\section{Mathematical formulations}
\label{sec:Mathematical_Formulations}

Consider an $N$-particle system evolving in dimension $d$ with position $\q_{i} \in \mathbb{R}^{d}$, momentum $\p_{i} \in \mathbb{R}^{d}$, and mass $m_{i} \in \mathbb{R}$ for $i=1,\dots,N$, the equations of motion for the non-Markovian form of the GLE are given by
\begin{subequations}\label{eq:GLE1}
\begin{align}
  \dd \q_{i} &= m_{i}^{-1}\p_{i} \dd t \, , \label{eq:GLE1-1} \\
  \dd \p_{i} &= - \nabla_{\q_{i}}U(\q) \dd t - \int^{t}_{0} \hat{K}(t-s) m_{i}^{-1}\p_{i}(s) \, \dd s \, \dd t + \etaB_{i} \dd t \, , \label{eq:GLE1-2}
\end{align}
\end{subequations}
where $U(\q)$ is a smooth potential energy. Component-wise, the random force $\eta^{x}_{i}(t)$, where the index $x$ indicates the appropriate Cartesian components, is a mean zero stationary Gaussian process with an autocorrelation function $\hat{K}$, satisfying the fluctuation-dissipation relation~\cite{Callen1951,Kubo1966},
\begin{equation}\label{eq:FDR}
  \langle \eta^{x}_{i}(t+s) \eta^{y}_{j}(t) \rangle = \beta^{-1} \hat{K}(s) \delta_{ij} \delta_{xy} \, , \quad s \geq 0 \, ,
\end{equation}
where $\beta$ denotes the inverse temperature, both $\delta_{ij}$ and $\delta_{xy}$ are Kronecker delta functions. In this article, we focus our attention on the following general form of the memory kernel:
\begin{equation}\label{eq:Kernel}
  \hat{K}(t) = \sum^{M}_{k=1} \tilde{\lambda}^{2}_{k} \exp\left( -\tilde{\alpha}_{k} t \right) \, , \quad t \geq 0 \, ,
\end{equation}
where $M$ is a positive integer that represent the number of modes,
\begin{equation}\label{eq:lambda_alpha_tilde}
  \tilde{\lambda}_{k} = \lambda_{k}/\sqrt{\epsilon} > 0 \, , \quad \tilde{\alpha}_{k} = \alpha_{k}/\epsilon > 0 \, ,
\end{equation}
with $\lambda_{k}$ and $\alpha_{k}$ being two constant parameters and $\epsilon > 0$ being a rescaling parameter (see more discussions in~\cite{Ottobre2011}). Note that for the sake of notational simplicity we do not explicitly express the dependence of $\tilde{\lambda}_{k}$ and $\tilde{\alpha}_{k}$ on $\epsilon$.

\subsection{Extended variable Markovian formalism}
\label{subsec:Markovian_Formalism}

In order to avoid dealing with the integral form of~\eqref{eq:GLE1-2}, it is desirable to rewrite~\eqref{eq:GLE1} as an extended variable Markovian formalism~\cite{Kupferman2004,Ceriotti2010,Ottobre2011,Baczewski2013,Pavliotis2014,Chak2020,Leimkuhler2020}. More specifically, following~\cite{Baczewski2013}, we define the extended variable, $\xiB_{i,k} \in \mathbb{R}^{d}$, associated with the $k$-th Prony mode's action on the $i$-th component of $\q$ and $\p$:
\begin{equation}\label{eq:xiB}
  \xiB_{i,k} = - \int^{t}_{0} \tilde{\lambda}_{k} \exp\left[ -\tilde{\alpha}_{k} (t-s) \right] m_{i}^{-1}\p_{i}(s) \, \dd s \, .
\end{equation}
Subsequently,~\eqref{eq:GLE1} may be rewritten as
\begin{subequations}\label{eq:GLE2}
\begin{align}
  \dd \q_{i} &= m_{i}^{-1}\p_{i} \dd t \, , \label{eq:GLE2-1} \\
  \dd \p_{i} &= - \nabla_{\q_{i}}U(\q) \dd t + \sum^{M}_{k=1} \tilde{\lambda}_{k} \xiB_{i,k} \dd t + \etaB_{i} \dd t \, . \label{eq:GLE2-2}
\end{align}
\end{subequations}
Differentiating~\eqref{eq:xiB} gives a simple stochastic differential equation (SDE):
\begin{equation}\label{eq:dxiB}
  \dd \xiB_{i,k} = - \tilde{\alpha}_{k} \xiB_{i,k} \dd t - \tilde{\lambda}_{k} m_{i}^{-1}\p_{i} \dd t \, .
\end{equation}
In order to construct a random force that satisfies the fluctuation-dissipation relation~\eqref{eq:FDR}, we consider the following SDE:
\begin{equation}\label{eq:detaB}
  \dd \etaB_{i,k} = - \tilde{\alpha}_{k} \etaB_{i,k} \dd t + \sqrt{ 2\tilde{\alpha}_{k} \beta^{-1} } \dd \WB_{i,k} \, ,
\end{equation}
where $\WB_{i,k} = \WB_{i,k}(t) \in \mathbb{R}^{d}$ is a vector of uncorrelated standard Wiener processes. It can be easily seen that~\eqref{eq:detaB} is an Ornstein--Uhlenbeck (OU) process where, component-wise, $\eta^{x}_{i,k}(t)$ has mean zero (i.e., $\langle \eta^{x}_{i,k} \rangle = 0$) and time correlation function~\cite{Shang2019} of
\begin{equation}
  \langle \eta^{x}_{i,k}(t+s) \eta^{x}_{i,k}(t) \rangle = \beta^{-1} \exp\left( -\tilde{\alpha}_{k} s \right) \, , \quad s \geq 0 \, .
\end{equation}
Therefore, the random force $\etaB_{i}$ in~\eqref{eq:GLE1-2} can be rewritten as
\begin{equation}\label{eq:etaB}
  \etaB_{i} = \sum^{M}_{k=1} \tilde{\lambda}_{k} \etaB_{i,k} \, .
\end{equation}
Moreover, combining the results of~\eqref{eq:Kernel}--\eqref{eq:etaB} and introducing a new variable of $\z_{i,k} = \xiB_{i,k} + \etaB_{i,k}$, the GLE~\eqref{eq:GLE1} may be rewritten as the following extended variable Markovian form:
\begin{subequations}\label{eq:GLE}
\begin{align}
  \dd \q_{i} &= m_{i}^{-1}\p_{i} \dd t \, , \label{eq:GLE-1} \\
  \dd \p_{i} &= - \nabla_{\q_{i}}U(\q) \dd t + \sum^{M}_{k=1} \tilde{\lambda}_{k} \z_{i,k} \dd t \, , \label{eq:GLE-2} \\
  \dd \z_{i,k} &= - \tilde{\lambda}_{k} m_{i}^{-1}\p_{i} \dd t - \tilde{\alpha}_{k} \z_{i,k} \dd t + \sqrt{ 2\tilde{\alpha}_{k} \beta^{-1} } \dd \WB_{i,k} \, , \quad k=1,\dots,M \, . \label{eq:GLE-3}
\end{align}
\end{subequations}
It is worth mentioning that, in the special case of $M=1$,~\eqref{eq:GLE} is very similar to a third-order Langevin dynamics that has certain advantages in controlling the discretization errors over its corresponding second-order form, i.e., an underdamped Langevin dynamics (see more discussions in~\cite{Mou2021}). Note also that, by introducing an additional variable, the extended variable Markovian form~\eqref{eq:GLE} can be cast into the GENERIC (General Equation for Non-Equilibrium Reversible--Irreversible Coupling) formalism (see more discussions in~\cite{Oettinger2005,Duong2013,Kraaij2020}), while this cannot be easily done in the original non-Markovian form~\eqref{eq:GLE1}. We can write down the Fokker--Planck (or forward Kolmogorov) equation associated with~\eqref{eq:GLE}
\begin{equation}\label{eq:FP}
\begin{aligned}
  \frac{ \partial \rho}{ \partial t} & \, =  \mathcal{L}^{\dag}_{\mathrm{GLE}} \rho = \sum^{N}_{i=1} \left[ - m_{i}^{-1}\p_{i} \cdot \nabla_{\q_{i}} \rho + \nabla_{\q_{i}}U(\q) \cdot \nabla_{\p_{i}} \rho \right] \\
  & + \sum^{N}_{i=1} \sum^{M}_{k=1} \left[ \tilde{\lambda}_{k} \left( -\z_{i,k} \cdot \nabla_{\p_{i}} \rho + m_{i}^{-1}\p_{i} \cdot \nabla_{\z_{i,k}} \rho \right) + \tilde{\alpha}_{k} \left( \nabla_{\z_{i,k}} \cdot \left( \z_{i,k} \rho \right) + \beta^{-1} \Delta_{\z_{i,k}} \rho \right) \right] \, .
\end{aligned}
\end{equation}
It can then be shown that there exists a unique invariant measure defined by the density
\begin{equation}\label{eq:rho_beta}
  \rho_{\beta}(\q,\p,\z) = \frac{1}{\hat{Z}} \exp \left( -\beta \left[ U(\q) + \sum^{N}_{i=1} \frac{ \p_{i} \cdot \p_{i} }{2m_{i}} + \sum^{N}_{i=1} \sum^{M}_{k=1} \frac{ \z_{i,k} \cdot \z_{i,k} }{2} \right] \right) \, ,
\end{equation}
where $\hat{Z}$ is the partition function. That is,~\eqref{eq:rho_beta} is the unique solution of the stationary Fokker--Planck equation~\eqref{eq:FP}, i.e., $\mathcal{L}^{\dag}_{\mathrm{GLE}} \rho_{\beta} = 0$.

\subsection{White noise limit}


From~\eqref{eq:GLE-3} we have
\begin{equation}
  \z_{i,k} \dd t = \frac{1}{\tilde{\alpha}_{k}} \left( - \dd \z_{i,k} - \tilde{\lambda}_{k} m_{i}^{-1}\p_{i} \dd t + \sqrt{ 2\tilde{\alpha}_{k} \beta^{-1} } \dd \WB_{i,k} \right) \, ,
\end{equation}
and substituting it into~\eqref{eq:GLE-2} gives
\begin{equation}
  \dd \p_{i} = - \nabla_{\q_{i}}U(\q) \dd t - \sum^{M}_{k=1} \frac{\tilde{\lambda}_{k}}{\tilde{\alpha}_{k}} \dd \z_{i,k} - \sum^{M}_{k=1} \frac{\tilde{\lambda}^{2}_{k}}{\tilde{\alpha}_{k}} m_{i}^{-1}\p_{i} \dd t + \sum^{M}_{k=1} \sqrt{ \frac{2\tilde{\lambda}^{2}_{k}\beta^{-1}}{\tilde{\alpha}_{k}} } \dd \WB_{i,k} \, ,
\end{equation}
which, in the white noise limit of $\epsilon \rightarrow 0$~\eqref{eq:lambda_alpha_tilde}, becomes
\begin{equation}
  \dd \p_{i} = - \nabla_{\q_{i}}U(\q) \dd t - \sum^{M}_{k=1} \frac{\lambda^{2}_{k}}{\alpha_{k}} m_{i}^{-1}\p_{i} \dd t + \sum^{M}_{k=1} \sqrt{ \frac{2\lambda^{2}_{k}\beta^{-1}}{\alpha_{k}} } \dd \WB_{i,k} \, .
\end{equation}
Rewriting the summation of $\dd \WB_{i,k}$ in the equation above as
\begin{equation}
  \sqrt{ \sum^{M}_{k=1} \frac{2\lambda^{2}_{k}\beta^{-1}}{\alpha_{k}} } \dd \WB_{i} \, ,
\end{equation}
we have
\begin{equation}
  \dd \p_{i} = - \nabla_{\q_{i}}U(\q) \dd t - \gamma m_{i}^{-1}\p_{i} \dd t + \sqrt{2\gamma\beta^{-1}} \dd \WB_{i} \, ,
\end{equation}
where the friction coefficient is given by
\begin{equation}
  \gamma = \sum^{M}_{k=1} \frac{\lambda^{2}_{k}}{\alpha_{k}} \, ,
\end{equation}
which is precisely the integral of the memory kernel defined in~\eqref{eq:Kernel},
\begin{equation}
  \gamma = \int^{\infty}_{0} \hat{K}(t) \, \dd t \, .
\end{equation}
Thus, we have heuristically verified that, in the white noise limit of $\epsilon \rightarrow 0$, the solution of the GLE~\eqref{eq:GLE} converges weakly to that of the Langevin dynamics (see a rigorous proof in~\cite{Ottobre2011}).

\subsection{Markovian approximations of the GLE with a more general memory kernel}
\label{subsec:Markovian_Approximations}

We have demonstrated in Section~\ref{subsec:Markovian_Formalism} that, for the particular form of the memory kernel~\eqref{eq:Kernel}, the GLE~\eqref{eq:GLE1} may be rewritten as the extended variable Markovian form~\eqref{eq:GLE}. We consider here the case of $N=d=1$ while dropping the subscripts for simplicity. Moreover, for a more general form of the memory kernel, we may be able to approximate the trajectories of $(q, p)$ in~\eqref{eq:GLE1} by trajectories of
$(\tilde{q}, \tilde{p})$ that solve a Markovian system with a vector of $M$ auxiliary variables, $\xiB=\left(\xi_1, \xi_2,\ldots,\xi_M\right)^\mathrm{T} \in \mathbb{R}^{M}$:
\begin{subequations}\label{eq:GLE3}
\begin{align}
  \dd \tilde{q} &= m^{-1} \tilde{p} \dd t \, , \label{eq:GLE3-1} \\
  \dd \tilde{p} &= - \nabla U(\tilde{q}) \dd t + \g^{\mathrm{T}} \xiB \dd t \, ,\\
  \dd \xiB&= - m^{-1} \tilde{p} \g \dd t - \A \xiB \dd t + \C \dd \WB \, , \quad \xiB(0)\sim \mathcal{N}\left(\mathbf{0},\SigmaB\right) \, ,
   \label{eq:GLE3-2}
\end{align}
\end{subequations}
where $\g \in \mathbb{R}^{M}$ is a constant vector, $\mathbf{A}, \mathbf{C} \in \mathbb{R}^{M \times M}$ are constant matrices, and $\WB = \WB(t) \in \mathbb{R}^{M}$ is a vector of uncorrelated standard Wiener processes. As demonstrated in~\cite{Kupferman2004},~\eqref{eq:GLE3} would be a Markovian approximation of~\eqref{eq:GLE1} if $\g, \A, \C$ and $\SigmaB$ satisfy the following relations:
\begin{equation}
\label{eq:relations}
  \SigmaB = \beta^{-1} \I \, , \quad \C\C^{\mathrm{T}} = \beta^{-1} \left(\A + \A^{\mathrm{T}}\right) \, ,
\end{equation}
with the approximated memory kernel being
\begin{equation}
  \tilde{K}(t) = \g^{\mathrm{T}} e^{-\A t} \g \, .
\end{equation}
We can also write down the Laplace transform of the approximated memory kernel as
\begin{equation}\label{eq:rational_function}
\mathcal{L}\{\tilde{K}\}(s) = \g^{\mathrm{T}} \left( \A + s \I \right)^{-1} \g \, ,
\end{equation}
where the right-hand side is a rational function of $s$. Thus, a Markovian approximation is established in two steps: first we approximate the Laplace transform of the memory kernel by a rational function; then we construct a matrix $\A$ and a vector $\g$ so that~\eqref{eq:rational_function} holds. Subsequently, $\SigmaB$ and $\C$ are determined from~\eqref{eq:relations}.

For a general memory kernel $\hat{K}$, these tasks are nontrivial. The simplest case is when the memory kernel can be approximated by a sum of exponentials, e.g., the positive Prony series memory kernel~\eqref{eq:Kernel}, whose Laplace transform can be easily computed as
\begin{equation}
  \quad \mathcal{L}\{\tilde{K}\}(s) = \sum_{k=1}^{M} \frac{\tilde{\lambda}^2_{k}}{s+\tilde{\alpha}_{k}} \, .
\end{equation}
In this case, $\A$ and $\g$ are chosen as
\begin{equation}
  \A = \mathrm{diag} \left(\tilde{\alpha}_1,\ldots, \tilde{\alpha}_M \right) \, , \quad \g = \left( \tilde{\lambda}_1,\ldots, \tilde{\lambda}_M \right)^{\mathrm{T}} \, .
\end{equation}
Note also that in cases where $\A$ is diagonalizable they can be reduced to the case above after an appropriate orthogonal transformation~\cite{Kupferman2004}.

\section{Numerical methods}
\label{sec:Numerical_Methods}

Splitting methods have been widely used in a range of systems, including Hamiltonian dynamics~\cite{Leimkuhler2005,Hairer2006}, dissipative systems~\cite{Shang2018}, and various stochastic dynamics~\cite{Leimkuhler2013,Leimkuhler2013a,Leimkuhler2013c,Leimkuhler2015,Leimkuhler2015a,Leimkuhler2016a,Shang2017,Shang2020}. The techniques have also been adopted in the construction of numerical methods for the GLE (e.g.,~\cite{Ceriotti2010,Baczewski2013,Leimkuhler2020}). In what follows we will first review three splitting methods proposed by Baczewski and Bond~\cite{Baczewski2013}. We will then propose a new method based on an alternative splitting of the vector field. Error analysis on the averages in the case of a one-dimensional harmonic oscillator will also be performed.

\subsection{The BACSCAB method}
\label{subsec:BACSCAB}

The vector field of the GLE~\eqref{eq:GLE} can be decomposed into pieces, for instance, A, B, C, and S:
\begin{equation}\label{eq:Splitting_ABCS}
\begin{aligned}
  \dd \left[ \begin{array}{c} \q_{i} \\ \p_{i} \\ \z_{i,1} \\ \z_{i,2} \\ \vdots \\ \z_{i,M} \end{array} \right] =& \, \underbrace{\left[ \begin{array}{c} m_{i}^{-1}\p_{i} \\ \mathbf{0} \\ \mathbf{0} \\ \mathbf{0} \\ \vdots \\ \mathbf{0} \end{array} \right] \dd t}_\mathrm{A} + \underbrace{\left[ \begin{array}{c} \mathbf{0} \\ - \nabla_{\q_{i}}U(\q) \\ \mathbf{0} \\ \mathbf{0} \\ \vdots \\ \mathbf{0} \end{array} \right] \dd t }_\mathrm{B} + \underbrace{\left[ \begin{array}{c} \mathbf{0} \\ \sum^{M}_{k=1} \tilde{\lambda}_{k} \z_{i,k} \\ \mathbf{0} \\ \mathbf{0} \\ \vdots \\ \mathbf{0} \end{array} \right] \dd t }_\mathrm{C} \\
  & + \underbrace{\left[ \begin{array}{c} \mathbf{0} \\ \mathbf{0} \\ - \tilde{\lambda}_{1} m_{i}^{-1}\p_{i} \dd t - \tilde{\alpha}_{1} \z_{i,1} \dd t + \sqrt{ 2\tilde{\alpha}_{1} \beta^{-1} } \dd \WB_{i,1} \\ - \tilde{\lambda}_{2} m_{i}^{-1}\p_{i} \dd t - \tilde{\alpha}_{2} \z_{i,2} \dd t + \sqrt{ 2\tilde{\alpha}_{2} \beta^{-1} } \dd \WB_{i,2} \\ \vdots \\ - \tilde{\lambda}_{M} m_{i}^{-1}\p_{i} \dd t - \tilde{\alpha}_{M} \z_{i,M} \dd t + \sqrt{ 2\tilde{\alpha}_{M} \beta^{-1} } \dd \WB_{i,M} \end{array} \right]}_\mathrm{S} \, ,
\end{aligned}
\end{equation}
in such a way that each subsystem can be solved ``exactly''. It is worth mentioning that the S part consists of a vector of uncorrelated OU processes, each of which has an exact (in the sense of distributional fidelity) solution~\cite{Kloeden1992}.

In describing splitting methods, we use the formal notation of the generator of the diffusion as in~\cite{DeFabritiis2006,Serrano2006,Thalmann2007}. The generators for each part of the system may be written out as follows:
\begin{subequations}
\begin{align}
  \mathcal{L}_{\mathrm{A}} &= \sum^{N}_{i=1} m_{i}^{-1}\p_{i} \cdot \nabla_{\mathbf{q}_{i}} \, , \label{eq:generator_A} \\
  \mathcal{L}_{\mathrm{B}} &= - \sum^{N}_{i=1} \nabla_{\mathbf{q}_{i}}U(\mathbf{q}) \cdot \nabla_{\mathbf{p}_{i}} \, , \label{eq:generator_B} \\
  \mathcal{L}_{\mathrm{C}} &= \sum^{N}_{i=1} \sum^{M}_{k=1}  \tilde{\lambda}_{k} \z_{i,k} \cdot \nabla_{\mathbf{p}_{i}} \, , \label{eq:generator_C} \\
  \mathcal{L}_{\mathrm{S}} &= \sum^{N}_{i=1} \sum^{M}_{k=1}  \left[ - \tilde{\lambda}_{k} m_{i}^{-1}\p_{i} - \tilde{\alpha}_{k} \z_{i,k} + \tilde{\alpha}_{k} \beta^{-1} \nabla_{\z_{i,k}} \right] \cdot \nabla_{\z_{i,k}} \, . \label{eq:generator_S}
\end{align}
\end{subequations}
The generator for the GLE thus can be written as
\begin{equation}\label{eq:generator_GLE}
  \mathcal{L}_{\mathrm{GLE}} = \mathcal{L}_{\mathrm{A}} + \mathcal{L}_{\mathrm{B}} + \mathcal{L}_{\mathrm{C}} + \mathcal{L}_{\mathrm{S}} \, .
\end{equation}
Moreover, the flow map (or phase space propagator) of the system may be given by the shorthand notation
\begin{equation}
  \mathcal{F}_{t} = e^{t \mathcal{L}_{\mathrm{GLE}}} \, ,
\end{equation}
where the exponential map is used to formally denote the solution operator to the equation $\partial_{t} u = \mathcal{L}_{\mathrm{GLE}} u$. Furthermore, approximations of $\mathcal{F}_{t}$ may be obtained as products (taken in different arrangements) of exponentials of the various splitting terms. For instance, the phase space propagation of a splitting method in Section II E of~\cite{Baczewski2013}, termed BACSCAB, can be written as
\begin{equation}\label{eq:Propagator_BACSCAB}
  e^{\Delta t\hat{\mathcal{L}}_\mathrm{BACSCAB}} = e^{\frac{\Delta t}{2}\mathcal{L}_\mathrm{B}} e^{\frac{\Delta t}{2}\mathcal{L}_\mathrm{A}} e^{\frac{\Delta t}{2}\mathcal{L}_\mathrm{C}} e^{\Delta t\mathcal{L}_\mathrm{S}} e^{\frac{\Delta t}{2}\mathcal{L}_\mathrm{C}} e^{\frac{\Delta t}{2}\mathcal{L}_\mathrm{A}} e^{\frac{\Delta t}{2}\mathcal{L}_\mathrm{B}} \, ,
\end{equation}
where $\exp\left(\Delta t\mathcal{L}_f\right)$ denotes the phase space propagator associated with the corresponding vector field $f$. Note that the steplengths associated with various operations are uniform and span the interval $\Delta t$. Therefore, each of the A, B, and C steps in~\eqref{eq:Propagator_BACSCAB} is taken with a steplength of $\Delta t/2$, while a steplength of $\Delta t$ is associated with the S step. Moreover, it is worth mentioning that, while each of the OU processes in the S step is exactly solvable, known as the ``method 2'' in~\cite{Baczewski2013},
\begin{equation}\label{eq:Exact_sols_z}
  \z_{i,k}(\Delta t) = \theta_{k} \z_{i,k}(0) - \left( 1 - \theta_{k} \right) \frac{ \tilde{\lambda}_{k} m_{i}^{-1}\p_{i} }{ \tilde{\alpha}_{k} } + \eta_{k} \sqrt{ \beta^{-1} } \RB_{i,k} \, ,
\end{equation}
where
\begin{equation}\label{eq:theta_eta}
  \theta_{k} = \exp\left(- \tilde{\alpha}_{k} \Delta t\right) \, , \quad \eta_{k} = \sqrt{ 1 - \theta^{2}_{k} } \, ,
\end{equation}
$\z_{i,k}(0)$ is the initial value of $\z_{i,k}$, and $\RB_{i,k}$ is a vector of independent and identically distributed (i.i.d.) standard normal random variables. Note that an alternative approach was used in the BACSCAB method, known as the ``method 3'' in~\cite{Baczewski2013}, by slightly modifying the $\eta_{k}$ defined above,
\begin{equation}\label{eq:eta_tilde}
  \tilde{\eta}_{k} = \sqrt{ \frac{2\left( 1 - \theta_{k} \right)^{2} }{ \Delta t \tilde{\alpha}_{k}} } \, ,
\end{equation}
in order to improve the stability~\cite{Baczewski2013}. However, we would like to point that in the current form there does not appear to have stability issues in the exact solutions~\eqref{eq:Exact_sols_z}--\eqref{eq:theta_eta} as $\tilde{\alpha}_{k} \rightarrow \infty$. (The memory kernel was written in a slightly different way in~\cite{Baczewski2013}.) The Euler--Maruyama method can also be used for solving the OU process, known as the ``method 1'' in~\cite{Baczewski2013}. However, it will not be included for comparisons since it has been observed that its performance is not as good as the other two alternative methods. The integration steps
of the BACSCAB method~\eqref{eq:Propagator_BACSCAB} read
\begin{subequations}
\begin{align}
  \p_{i}^{n+1/4} & = \p_{i}^{n} - (\Delta t/2) \nabla_{\q_{i}^{n}}U(\q^{n}) \, , \\
  \q_{i}^{n+1/2} & = \q_{i}^{n} + (\Delta t/2) m_{i}^{-1} \p_{i}^{n+1/4} \, , \\
  \p_{i}^{n+2/4} & = \p_{i}^{n+1/4} + (\Delta t/2) \sum^{M}_{k=1} \tilde{\lambda}_{k} \z_{i,k}^{n} \, , \\
  \z_{i,k}^{n+1} &= \theta_{k} \z_{i,k}^{n} - \left( 1 - \theta_{k} \right) \frac{ \tilde{\lambda}_{k} m_{i}^{-1}\p_{i}^{n+2/4} }{ \tilde{\alpha}_{k} } + \tilde{\eta}_{k} \sqrt{ \beta^{-1} } \RB_{i,k}^{n} \, , \\
  \p_{i}^{n+3/4} & = \p_{i}^{n+2/4} + (\Delta t/2) \sum^{M}_{k=1} \tilde{\lambda}_{k} \z_{i,k}^{n+1} \, , \\
  \q_{i}^{n+1} & = \q_{i}^{n+1/2} + (\Delta t/2) m_{i}^{-1} \p_{i}^{n+3/4} \, , \\
  \p_{i}^{n+1} & = \p_{i}^{n+3/4} - (\Delta t/2) \nabla_{\q_{i}^{n+1}}U(\q^{n+1}) \, .
\end{align}
\end{subequations}

\subsection{The PASP method}
\label{subsec:PASP}

An alternative splitting method was proposed in Section II C of~\cite{Baczewski2013}, termed PASP, whose phase space propagation can be written as
\begin{equation}\label{eq:Propagator_PASP}
  \exp\left(\Delta t \hat{\mathcal{L}}_\mathrm{PASP} \right) =
  \exp\left(\frac{\Delta t}{2}\mathcal{L}_\mathrm{P}\right) \exp\left(\Delta t \mathcal{L}_\mathrm{A}\right)
  \exp\left(\Delta t \mathcal{L}_\mathrm{S}\right)
  \exp\left(\frac{\Delta t}{2} \mathcal{L}_\mathrm{P}\right) \, ,
\end{equation}
where
\begin{equation}\label{eq:generator_P}
  \mathcal{L}_\mathrm{P} = \mathcal{L}_\mathrm{B} + \mathcal{L}_\mathrm{C} \, .
\end{equation}
Depending on how the S part is solved, there are three variants of the PASP method. For example, if the S part is solved exactly as in~\eqref{eq:Exact_sols_z}--\eqref{eq:theta_eta}, we arrive the PASP-2 method, whose integration steps read
\begin{subequations}
\begin{align}
  \p_{i}^{n+1/2} & = \p_{i}^{n} - (\Delta t/2) \nabla_{\q_{i}^{n}}U(\q^{n}) + (\Delta t/2) \sum^{M}_{k=1} \tilde{\lambda}_{k} \z_{i,k}^{n} \, , \label{eq:PASP-1} \\
  \q_{i}^{n+1} & = \q_{i}^{n} + \Delta t m_{i}^{-1} \p_{i}^{n+1/2} \, , \label{eq:PASP-2} \\
  \z_{i,k}^{n+1} &= \theta_{k} \z_{i,k}^{n} - \left( 1 - \theta_{k} \right) \frac{ \tilde{\lambda}_{k} m_{i}^{-1}\p_{i}^{n+1/2} }{ \tilde{\alpha}_{k} } + \eta_{k} \sqrt{ \beta^{-1} } \RB_{i,k}^{n} \, , \label{eq:PASP-3} \\
  \p_{i}^{n+1} & = \p_{i}^{n+1/2} - (\Delta t/2) \nabla_{\q_{i}^{n+1}}U(\q^{n+1}) + (\Delta t/2) \sum^{M}_{k=1} \tilde{\lambda}_{k} \z_{i,k}^{n+1} \, . \label{eq:PASP-4}
\end{align}
\end{subequations}
Note that the integration steps for the PASP-3 method~\cite{Baczewski2013} are exactly the same as PASP-2 except the $\eta_{k}$ in~\eqref{eq:PASP-3} is replaced by $\tilde{\eta}_{k}$ as in~\eqref{eq:eta_tilde}.

\subsection{The BAEOEAB method}
\label{subsec:BAEOEAB}

BACSCAB and both PASP methods mentioned above rely on solving the S part exactly as in~\eqref{eq:Exact_sols_z}--\eqref{eq:theta_eta} or with a slight modification~\eqref{eq:eta_tilde}. However, one potential drawback of that approach is, as $\tilde{\alpha}_{k} \rightarrow \infty$, the update on $\z_{i,k}$ becomes increasingly dominated by the noise in~\eqref{eq:Exact_sols_z}, thereby shrinking the contribution from the ``force term'' (i.e., $- \tilde{\lambda}_{k} m_{k}^{-1}\p_{k}$), which is likely to cause potential accuracy and/or stability issues in sampling the invariant measure~\eqref{eq:rho_beta}. To this end, we propose to combine the ``force term'' with the original C part in~\eqref{eq:Splitting_ABCS} to form the E part as follows:
\begin{equation}\label{eq:Splitting_ABEO}
\begin{aligned}
  \dd \left[ \begin{array}{c} \q_{i} \\ \p_{i} \\ \z_{i,1} \\ \z_{i,2} \\ \vdots \\ \z_{i,M} \end{array} \right] =& \, \underbrace{\left[ \begin{array}{c} m_{i}^{-1}\p_{i} \\ \mathbf{0} \\ \mathbf{0} \\ \mathbf{0} \\ \vdots \\ \mathbf{0} \end{array} \right] \dd t}_\mathrm{A} + \underbrace{\left[ \begin{array}{c} \mathbf{0} \\ - \nabla_{\q_{i}}U(\q) \\ \mathbf{0} \\ \mathbf{0} \\ \vdots \\ \mathbf{0} \end{array} \right] \dd t }_\mathrm{B} + \underbrace{\left[ \begin{array}{c} \mathbf{0} \\ \sum^{M}_{k=1} \tilde{\lambda}_{k} \z_{i,k} \\ - \tilde{\lambda}_{1} m_{i}^{-1}\p_{i} \\ - \tilde{\lambda}_{2} m_{i}^{-1}\p_{i} \\ \vdots \\ - \tilde{\lambda}_{M} m_{i}^{-1}\p_{i} \end{array} \right] \dd t }_\mathrm{E} \\
  & + \underbrace{\left[ \begin{array}{c} \mathbf{0} \\ \mathbf{0} \\ - \tilde{\alpha}_{1} \z_{i,1} \dd t + \sqrt{ 2\tilde{\alpha}_{1} \beta^{-1} } \dd \WB_{i,1} \\ - \tilde{\alpha}_{2} \z_{i,2} \dd t + \sqrt{ 2\tilde{\alpha}_{2} \beta^{-1} } \dd \WB_{i,2} \\ \vdots \\ - \tilde{\alpha}_{M} \z_{i,M} \dd t + \sqrt{ 2\tilde{\alpha}_{M} \beta^{-1} } \dd \WB_{i,M} \end{array} \right]}_\mathrm{O} \, .
\end{aligned}
\end{equation}
In this case, each of the OU processes in the O step is still exactly solvable,
\begin{equation}\label{eq:Exact_sols_z_short}
  \z_{i,k}(\Delta t) = \theta_{k} \z_{i,k}(0) + \sqrt{ \beta^{-1} \left( 1 - \theta^{2}_{k} \right) } \RB_{i,k} \, .
\end{equation}
Moreover, we can further split the E part in such a way that each subsystem, $\mathrm{E^{x}_{i,k}}$, component-wise,
\begin{equation}\label{eq:Splitting_E}
  \dd \left[ \begin{array}{c} q_{i}^{x} \\ p_{i}^{x} \\ z_{i,k}^{x} \end{array} \right] = \left[ \begin{array}{c} 0 \\ \tilde{\lambda}_{k} z_{i,k}^{x} \\ - \tilde{\lambda}_{k} m_{i}^{-1}p_{i}^{x} \end{array} \right] \dd t \, ,
\end{equation}
with $q_{i}^{x}$ remaining fixed, corresponds to a harmonic oscillator with the exact solution:
\begin{subequations}\label{eq:Exact_sols_E}
\begin{align}
  p_{i}^{x}(\Delta t) &= \cos(\tilde{\lambda}_{k} m^{-1/2}_{i} \Delta t) p_{i}^{x}(0) + \sin(\tilde{\lambda}_{k} m^{-1/2}_{i} \Delta t) m^{1/2}_{i} z_{i,k}^{x}(0) \, , \\
  z_{i,k}^{x}(\Delta t) &= - \sin(\tilde{\lambda}_{k} m^{-1/2}_{i} \Delta t) m^{-1/2}_{i} p_{i}^{x}(0) + \cos(\tilde{\lambda}_{k} m^{-1/2}_{i} \Delta t) z_{i,k}^{x}(0) \, ,
\end{align}
\end{subequations}
where the superscript $x \leq d$ is a positive integer that represents a specific dimension, $p_{i}^{x}(0)$ and $z_{i,k}^{x}(0)$ are the initial values of $p_{i}^{x}$ and $z_{i,k}^{x}$, respectively. We further propose a new splitting method, termed BAEOEAB, whose phase space propagation can be written as
\begin{equation}\label{eq:Propagator_BAEOEAB}
  e^{\Delta t\hat{\mathcal{L}}_\mathrm{BAEOEAB}} = e^{\frac{\Delta t}{2}\mathcal{L}_\mathrm{B}} e^{\frac{\Delta t}{2}\mathcal{L}_\mathrm{A}} e^{\frac{\Delta t}{2}\hat{\mathcal{L}}_\mathrm{E}} e^{\Delta t\mathcal{L}_\mathrm{O}} e^{\frac{\Delta t}{2}\hat{\mathcal{L}}_\mathrm{E}} e^{\frac{\Delta t}{2}\mathcal{L}_\mathrm{A}} e^{\frac{\Delta t}{2}\mathcal{L}_\mathrm{B}} \, .
\end{equation}
Given that the E part has been further split, the propagation of either of the E parts in~\eqref{eq:Propagator_BAEOEAB} should be more explicitly defined as
\begin{equation}\label{eq:Propagator_E}
  e^{\frac{\Delta t}{2}\hat{\mathcal{L}}_\mathrm{E}} = e^{\frac{\Delta t}{2}\mathcal{L}_{\mathrm{E}^{z}_{N,M}}}
  e^{\frac{\Delta t}{2}\mathcal{L}_{\mathrm{E}^{y}_{N,M}}}
  e^{\frac{\Delta t}{2}\mathcal{L}_{\mathrm{E}^{x}_{N,M}}}
  \dots
  e^{\frac{\Delta t}{2}\mathcal{L}_{\mathrm{E}^{z}_{1,1}}}
  e^{\frac{\Delta t}{2}\mathcal{L}_{\mathrm{E}^{y}_{1,1}}}
  e^{\frac{\Delta t}{2}\mathcal{L}_{\mathrm{E}^{x}_{1,1}}}
   \, .
\end{equation}
Note that one may wish to reverse the order in either of the E parts in~\eqref{eq:Propagator_BAEOEAB}, which would affect neither its overall performance nor the order of convergence to the invariant measure. The corresponding integration steps may be written out
as follows:
\begin{subequations}
\begin{align}
  \p_{i}^{n+1/4} & = \p_{i}^{n} - (\Delta t/2) \nabla_{\q_{i}^{n}}U(\q^{n}) \, , \\
  \q_{i}^{n+1/2} & = \q_{i}^{n} + (\Delta t/2) m_{i}^{-1} \p_{i}^{n+1/4} \, , \\
  p_{i}^{x,n+2/4} &= \cos(\tilde{\lambda}_{k} m^{-1/2}_{i} \Delta t/2) p_{i}^{x,n+1/4} + \sin(\tilde{\lambda}_{k} m^{-1/2}_{i} \Delta t/2) m^{1/2}_{i} z_{i,k}^{x,n} \, , \label{eq:BAEOEAB_c} \\
  z_{i,k}^{x,n+1/3} &= - \sin(\tilde{\lambda}_{k} m^{-1/2}_{i} \Delta t/2) m^{-1/2}_{i} p_{i}^{x,n+1/4} + \cos(\tilde{\lambda}_{k} m^{-1/2}_{i} \Delta t/2) z_{i,k}^{x,n} \, , \label{eq:BAEOEAB_d} \\
  \z_{i,k}^{n+2/3} &= \theta_{k} \z_{i,k}^{n+1/3} + \sqrt{ \beta^{-1} \left( 1 - \theta^{2}_{k} \right) } \RB_{i,k} \, , \\
  p_{i}^{x,n+3/4} &= \cos(\tilde{\lambda}_{k} m^{-1/2}_{i} \Delta t/2) p_{i}^{x,n+2/4} + \sin(\tilde{\lambda}_{k} m^{-1/2}_{i} \Delta t/2) m^{1/2}_{i} z_{i,k}^{x,n+2/3} \, , \label{eq:BAEOEAB_f} \\
  z_{i,k}^{x,n+1} &= - \sin(\tilde{\lambda}_{k} m^{-1/2}_{i} \Delta t/2) m^{-1/2}_{i} p_{i}^{x,n+2/4} + \cos(\tilde{\lambda}_{k} m^{-1/2}_{i} \Delta t/2) z_{i,k}^{x,n+2/3} \, , \label{eq:BAEOEAB_g} \\
  \q_{i}^{n+1} & = \q_{i}^{n+1/2} + (\Delta t/2) m_{i}^{-1} \p_{i}^{n+3/4} \, , \\
  \p_{i}^{n+1} & = \p_{i}^{n+3/4} - (\Delta t/2) \nabla_{\q_{i}^{n+1}}U(\q^{n+1}) \, .
\end{align}
\end{subequations}
As indicated in~\eqref{eq:Propagator_E}, both~\eqref{eq:BAEOEAB_c}--\eqref{eq:BAEOEAB_d} and ~\eqref{eq:BAEOEAB_f}--\eqref{eq:BAEOEAB_g} loop over not only each pair of $i=1\dots N$ and $k=1\dots M$, but also each Cartesian component (i.e., $x,y,$ or $z$). Note that we can combine the E and O parts in~\eqref{eq:Splitting_ABEO} together to form an OU process that can also be solved exactly as in~\cite{Ceriotti2010,Leimkuhler2020}. However, this approach relies on operations involving potentially very large matrices, which could be computationally very demanding (especially when the number of modes, $M$, is large and/or some of the parameters, $\tilde{\lambda}_{k}$ and $\tilde{\alpha}_{k}$, are, for instance, position-dependent) and thus we would like to avoid. It is also worth mentioning that we can easily adopt the procedures based on the framework of the long-time Talay--Tubaro expansion~\cite{Talay1990,Debussche2012,Leimkuhler2013,Leimkuhler2013a,Leimkuhler2013c,Abdulle2014,Abdulle2014a,Leimkuhler2015a,Leimkuhler2015b,Shang2020} to analyze the accuracy of ergodic averages (with respect to the invariant measure) in the stochastic numerical methods previously mentioned in this section (with a general nonlinear force), and conclude that they all have second order convergence to the invariant measure.

\subsection{Error analysis}
\label{subsec:Error_Analysis}

As in~\cite{Baczewski2013}, we adopt a standard test case typically used in classical Langevin dynamics~\cite{Leimkuhler2013a} by applying a one-dimensional harmonic oscillator $U(q) = Kq^{2}/2$ ($K > 0$) with a single mode (i.e., $M=1$ in~\eqref{eq:Kernel} while dropping the subscripts for simplicity to have $m$, $\tilde{\lambda}$, and $\tilde{\alpha}$).
In such a simple case, we can explicitly write down one iteration of a general numerical method evolving the dynamics as
\begin{equation}
\left[
  \begin{array}{c}
    q_{n+1} \\
    p_{n+1} \\
    z_{n+1} \\
  \end{array}
\right]
 =
\Psi
\left[
  \begin{array}{c}
    q_{n} \\
    p_{n} \\
    z_{n} \\
  \end{array}
\right]
+ \mu_{n} \, ,
\end{equation}
where $\Psi = (\psi_{ij})$, $i,j \in \{1,2,3\}$, is a constant matrix and $\mu_{n}$ is a vector of stochastic processes, whose components can be denoted as $\mu_{n,i}$.
Taking products of $q_{n+1}$, $p_{n+1}$, and $z_{n+1}$ in the update equations above
and then taking expectations on both sides of the equations with respect to the invariant measure of the numerical method yields
\begin{align*}
  \langle q^{2} \rangle = & \, \psi^{2}_{11} \langle q^{2} \rangle + \psi^{2}_{12} \langle p^{2} \rangle + \psi^{2}_{13} \langle z^{2} \rangle + \langle \mu^{2}_{1} \rangle + 2\psi_{11}\psi_{12} \langle qp \rangle + 2\psi_{11}\psi_{13} \langle qz \rangle + 2\psi_{12}\psi_{13} \langle pz \rangle \, , \\
  \langle p^{2} \rangle = & \, \psi^{2}_{21} \langle q^{2} \rangle + \psi^{2}_{22} \langle p^{2} \rangle + \psi^{2}_{23} \langle z^{2} \rangle + \langle \mu^{2}_{2} \rangle + 2\psi_{21}\psi_{22} \langle qp \rangle + 2\psi_{21}\psi_{23} \langle qz \rangle + 2\psi_{22}\psi_{23} \langle pz \rangle \, , \\
  \langle z^{2} \rangle = & \, \psi^{2}_{31} \langle q^{2} \rangle + \psi^{2}_{32} \langle p^{2} \rangle + \psi^{2}_{33} \langle z^{2} \rangle + \langle \mu^{2}_{3} \rangle + 2\psi_{31}\psi_{32} \langle qp \rangle + 2\psi_{31}\psi_{33} \langle qz \rangle + 2\psi_{32}\psi_{33} \langle pz \rangle \, , \\
  \langle qp \rangle = & \, \psi_{11}\psi_{21} \langle q^{2} \rangle + \psi_{12}\psi_{22} \langle p^{2} \rangle + \psi_{13}\psi_{23} \langle z^{2} \rangle + \langle \mu_{1}\mu_{2} \rangle + \left( \psi_{11}\psi_{22} + \psi_{12}\psi_{21} \right) \langle qp \rangle \\
  & + \left( \psi_{11}\psi_{23} + \psi_{13}\psi_{21} \right) \langle qz \rangle + \left( \psi_{12}\psi_{23} + \psi_{13}\psi_{22} \right) \langle pz \rangle \, , \\
  \langle qz \rangle = & \, \psi_{11}\psi_{31} \langle q^{2} \rangle + \psi_{12}\psi_{32} \langle p^{2} \rangle + \psi_{13}\psi_{33} \langle z^{2} \rangle + \langle \mu_{1}\mu_{3} \rangle + \left( \psi_{11}\psi_{32} + \psi_{12}\psi_{31} \right) \langle qp \rangle \\
  & + \left( \psi_{11}\psi_{33} + \psi_{13}\psi_{31} \right) \langle qz \rangle + \left( \psi_{12}\psi_{33} + \psi_{13}\psi_{32} \right) \langle pz \rangle \, , \\
  \langle pz \rangle = & \, \psi_{21}\psi_{31} \langle q^{2} \rangle + \psi_{22}\psi_{32} \langle p^{2} \rangle + \psi_{23}\psi_{33} \langle z^{2} \rangle + \langle \mu_{2}\mu_{3} \rangle + \left( \psi_{21}\psi_{32} + \psi_{22}\psi_{31} \right) \langle qp \rangle \\
  & + \left( \psi_{21}\psi_{33} + \psi_{23}\psi_{31} \right) \langle qz \rangle + \left( \psi_{22}\psi_{33} + \psi_{23}\psi_{32} \right) \langle pz \rangle \, ,
\end{align*}
where $\langle \cdot \rangle$ denotes an ensemble average with respect to the invariant measure of the numerical method. See examples in the case of Langevin dynamics in Chapter 7 of~\cite{Leimkuhler2015b}. While the analytical expressions to the averages on the left-hand side above are given respectively by
\begin{equation}
  \langle q^{2} \rangle = \frac{1}{K \beta} \, , \quad \langle p^{2} \rangle = \frac{m}{\beta} \, , \quad \langle z^{2} \rangle = \frac{1}{\beta} \, , \quad \langle qp \rangle = 0 \, , \quad \langle qz \rangle = 0 \, , \quad \langle pz \rangle = 0 \, ,
\end{equation}
we can also solve the above linear system to obtain the averages associated with each numerical method. For brevity, some results are shown as leading order series in $\Delta t$.

\subsubsection{PASP-2}


In the PASP-2 method~\eqref{eq:Propagator_PASP}, we have the following averages:
\begin{subequations}
\begin{align}
  \langle q^{2} \rangle
  &= \frac{1}{K \beta} \left[ 1 + \frac{\Delta t^{2}\left( m\tilde{\alpha}^{2} + 3K \right)}{12m} \right] + O(\Delta t^{4}) \, , \\
  \langle p^{2} \rangle &=
  \frac{m}{\beta} \left[ 1 + \frac{\Delta t^{2}\tilde{\alpha}^{2}}{12} \right] + O(\Delta t^{4}) \, , \\
  \langle z^{2} \rangle
  &= \frac{1}{\beta} \left[ 1 + \frac{\Delta t^{2}\tilde{\lambda}^{2}}{4m} \right] + O(\Delta t^{4}) \, , \\
  \langle qp \rangle &= \langle pz \rangle = 0 \, , \\
  \langle qz \rangle &=
  - \frac{\Delta t^{2}\tilde{\lambda}}{4m\beta} + O(\Delta t^{4}) \, .
\end{align}
\end{subequations}
While the averages of $\langle qp \rangle$ and $\langle pz \rangle$ are exact, there exist second order errors in the averages of $\langle q^{2} \rangle$, $\langle p^{2} \rangle$, $\langle z^{2} \rangle$, and $\langle qz \rangle$. Note also that the minus sign of the leading order term in $\langle qz \rangle$ appeared to be missing in~\cite{Baczewski2013}.

\subsubsection{PASP-3}


Similarly in the PASP-3 method, we have the following averages:
\begin{subequations}
\begin{align}
  \langle q^{2} \rangle
  &= \frac{1}{K \beta} \left[ 1 + \frac{\Delta t^{2}K}{4m} \right] + O(\Delta t^{4}) \, , \\
  \langle p^{2} \rangle &= \frac{m}{\beta} \, , \\
  \langle z^{2} \rangle
  &= \frac{1}{\beta} \left[ 1 + \frac{\Delta t^{2}\left( 3\tilde{\lambda}^{2} - m\tilde{\alpha}^{2} \right)}{12m} \right] + O(\Delta t^{4}) \, , \\
  \langle qp \rangle &= \langle pz \rangle = 0 \, , \\
  \langle qz \rangle &=
  - \frac{\Delta t^{2}\tilde{\lambda}}{4m\beta} + O(\Delta t^{4}) \, .
\end{align}
\end{subequations}
Similar to the case of the PASP-2 method, the averages of $\langle qp \rangle$ and $\langle pz \rangle$ are exact, while there exist second order errors in the averages of $\langle q^{2} \rangle$, $\langle p^{2} \rangle$, $\langle z^{2} \rangle$, and $\langle qz \rangle$. Again, the minus sign of the leading order term in $\langle qz \rangle$ appeared to be missing in~\cite{Baczewski2013}.

\subsubsection{BACSCAB}


In the BACSCAB method~\eqref{eq:Propagator_BACSCAB}, we have the following averages:
\begin{subequations}
\begin{align}
  \langle q^{2} \rangle &= \frac{1}{K \beta} \, , \\
  \langle p^{2} \rangle &= \frac{m}{\beta} \left[ 1 - \frac{\Delta t^{2}K}{4m} \right] \, , \\
  \langle z^{2} \rangle
  &= \frac{1}{\beta} \left[ 1 + \frac{\Delta t^{2}\left( 3\tilde{\lambda}^{2} - m\tilde{\alpha}^{2} \right)}{12m} \right] + O(\Delta t^{4}) \, , \\
  \langle qp \rangle &= \langle qz \rangle = \langle pz \rangle = 0 \, .
\end{align}
\end{subequations}
While the averages of $\langle q^{2} \rangle$, $\langle qp \rangle$, $\langle qz \rangle$, and $\langle pz \rangle$ are exact, there exist second order errors in the averages of $\langle p^{2} \rangle$ and $\langle z^{2} \rangle$.

\subsubsection{BAEOEAB}

In the BAEOEAB method~\eqref{eq:Propagator_BAEOEAB}, we have the following averages:
\begin{equation}
  \langle q^{2} \rangle = \frac{1}{K \beta} \, , \quad \langle p^{2} \rangle = \frac{m}{\beta} \left[ 1 - \frac{\Delta t^{2}K}{4m} \right] \, , \quad \langle z^{2} \rangle = \frac{1}{\beta} \, , \quad \langle qp \rangle = \langle qz \rangle = \langle pz \rangle = 0 \, .
\end{equation}
That is, all of the averages are exact, with the only exception of $\langle p^{2} \rangle$ that has second order errors, which yields a friction-independent upper bound of the stepsize $\Delta t_{\mathrm{max}}=2\sqrt{m/K}$, coinciding with the (deterministic) Verlet stability threshold~\cite{Leimkuhler2013a}. Note that a similar method to BAEOEAB (i.e., BAOEOAB) yields exactly the same averages, but requires additional generations of random numbers. It is also worth mentioning that the averages of $\langle q^{2} \rangle$, $\langle p^{2} \rangle$, and $\langle qp \rangle$ match perfectly with their counterparts in the BAOAB method of the underdamped Langevin dynamics~\cite{Leimkuhler2013a}.

\section{Numerical experiments}
\label{sec:Numerical_Experiments}

In this section, a variety of numerical experiments are conducted to systematically compare the newly proposed BAEOEAB method~\eqref{eq:Propagator_BAEOEAB} with alternative methods described in Section~\ref{sec:Numerical_Methods}. While all the methods presented can be used in multi-mode systems, we focus our attention on a single mode (i.e., $M=1$) for simplicity where the subscripts in $\tilde{\lambda}$ and $\tilde{\alpha}$~\eqref{eq:Kernel} can be dropped.

\subsection{Harmonic oscillator}

In order to verify the error analysis results in Section~\ref{subsec:Error_Analysis}, we compare the performance of various methods with a one-dimensional harmonic oscillator $U(q) = Kq^{2}/2$ ($K > 0$). The following set of parameters were used: $m=K=\beta=\tilde{\alpha}=1$ and $\tilde{\lambda}=2$.

\begin{figure}[tb]
\centering
\includegraphics[scale=0.45]{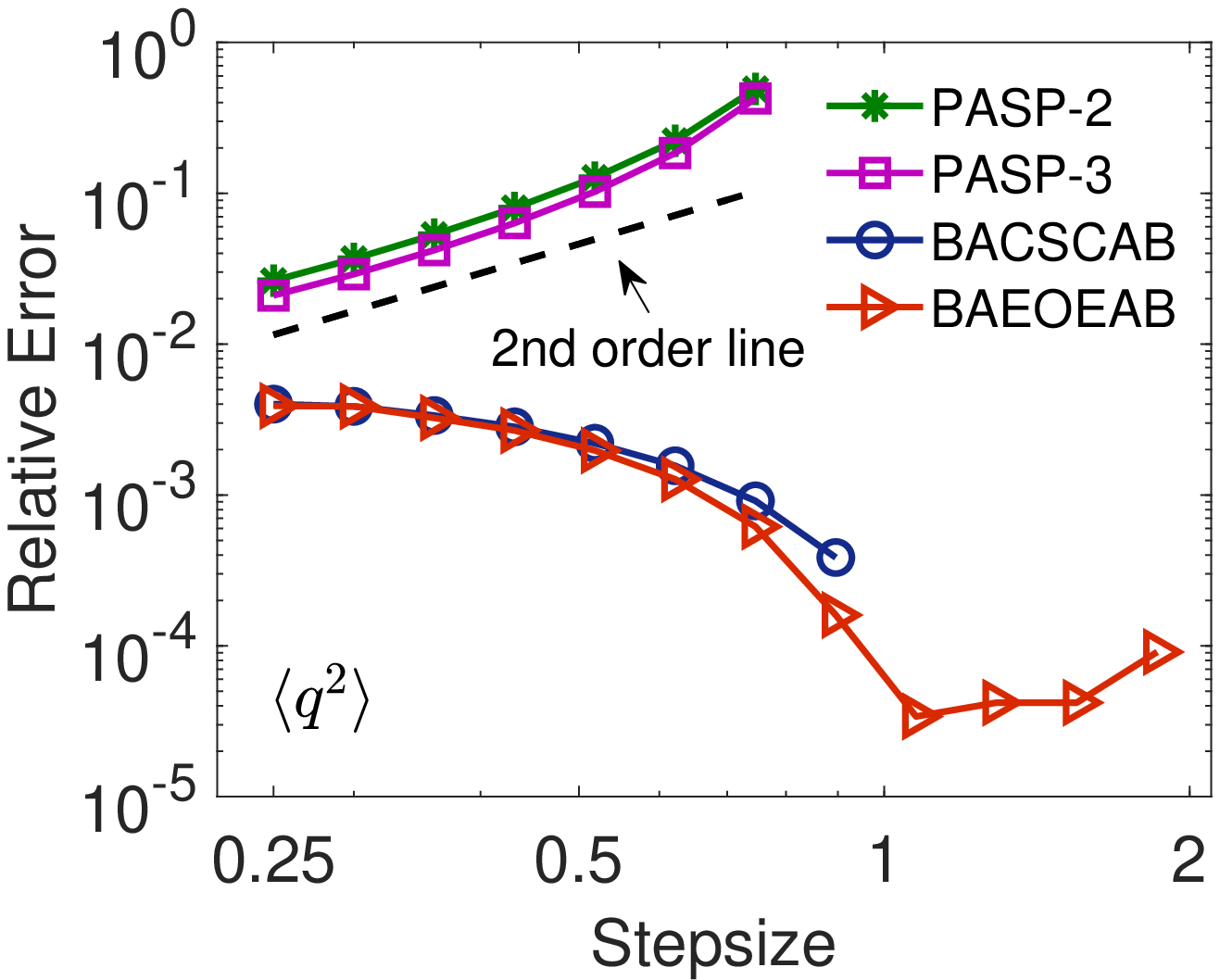}
\includegraphics[scale=0.45]{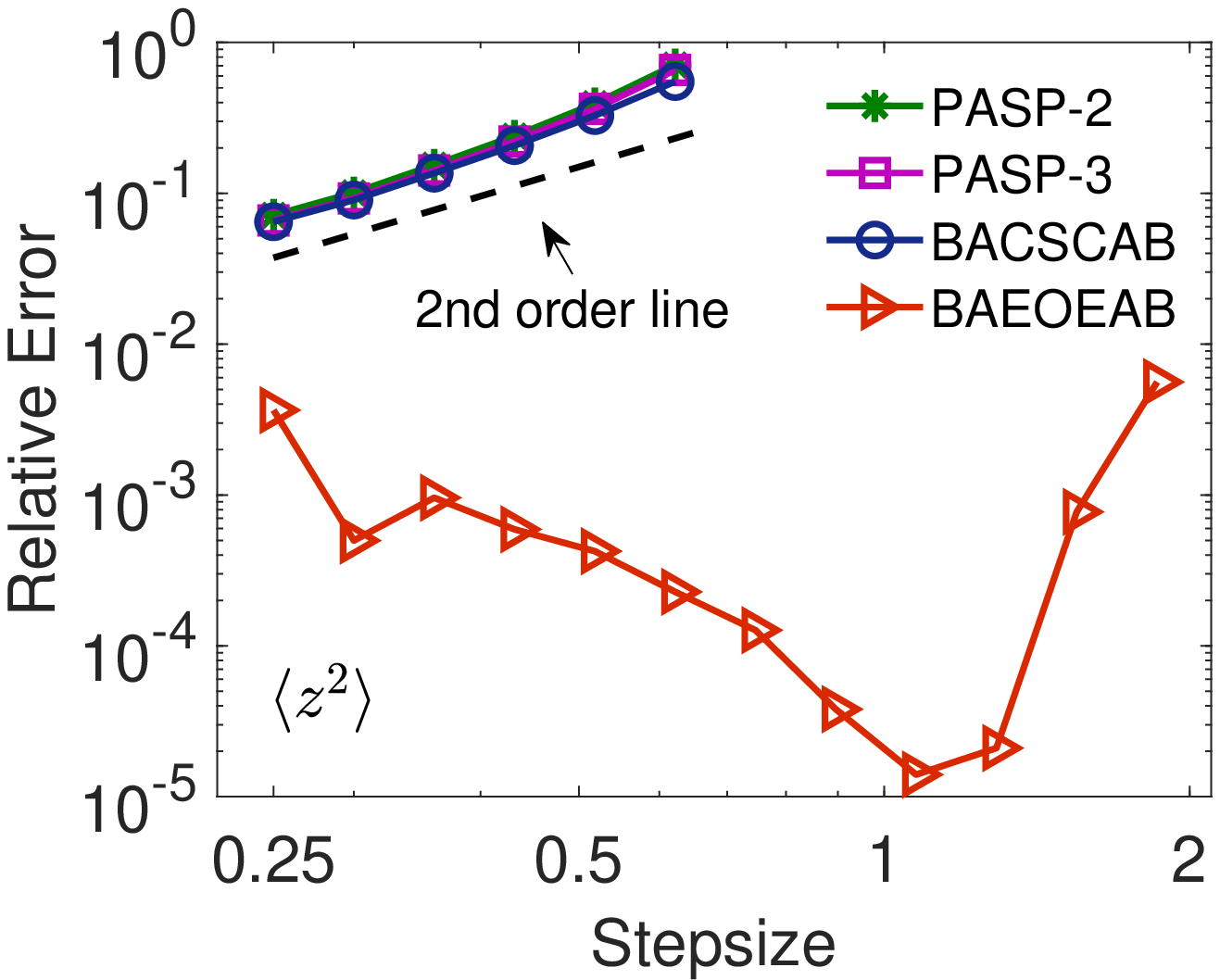}
\caption{\small Double logarithmic plot of the relative error in the computed averages of $\langle q^2 \rangle$ (left) and $\langle z^2 \rangle$ (right) against stepsize by using various numerical methods for the GLE described in Section~\ref{sec:Numerical_Methods} with parameters of $\tilde{\lambda}=2$ and $\tilde{\alpha}=1$. The system was simulated for $10^{7}$ reduced time units but only the last 80\% of the data were collected to calculate the static quantity in order to make sure the system was well equilibrated. 1000 different runs were averaged to reduce the sampling errors. The stepsizes tested began at $\Delta t=0.25$ and were increased incrementally by 20\% until all methods either started to show significant relative errors or became unstable. The dashed black line represents the second order convergence to the invariant measure.}
\label{fig:GLE_HO_Comp}
\end{figure}

Figure~\ref{fig:GLE_HO_Comp} compares the control of the relative errors in the computed averages of $\langle q^2 \rangle$ and $\langle z^2 \rangle$. According to the dashed order lines, both PSAP-2 and PASP-3 methods exhibit second order convergence in both averages whereas the BACSCAB method appears to be second order only in the average of $\langle z^2 \rangle$ on the right panel. Moreover, the BAEOEAB method is exact in both averages while the BACSCAB method is exact only in the average of $\langle q^2 \rangle$ on the left panel. It is worth mentioning that the error in cases where the corresponding averages are exact comes solely from the sampling error, rather than the discretization error. Note also that all the observations are consistent with our findings in Section~\ref{subsec:Error_Analysis}.

More precisely, for the average of $\langle q^2 \rangle$ on the left panel of Figure~\ref{fig:GLE_HO_Comp}, PASP-3 appears to be only slightly more accurate than PASP-2 while both methods are significantly outperformed by either BACSCAB or BAEOEAB in terms of accuracy. It is worth pointing out that the BAEOEAB method also appears to be more robust than alternative methods in a wide range of stepsizes---while both PASP-2 and PASP-3 methods exhibit around 50\% relative errors at a stepsize of around $\Delta t = 0.75$ and the BACSCAB method became unstable just over $\Delta t = 1$, the BAEOEAB method is still extremely accurate (i.e., up to sampling error) when the stepsize is around $\Delta t = 1.9$, close to its stability threshold $\Delta t_{\mathrm{max}}=2$. The behavior is largely similar for the average of $\langle z^2 \rangle$ on the right panel of Figure~\ref{fig:GLE_HO_Comp} except the performance of PASP-2, PASP-3, and BACSCAB are almost indistinguishable. In this case, the BAEOEAB method substantially outperforms alternative methods in terms of not only accuracy but also robustness.

\begin{figure}[tb]
\centering
\includegraphics[scale=0.45]{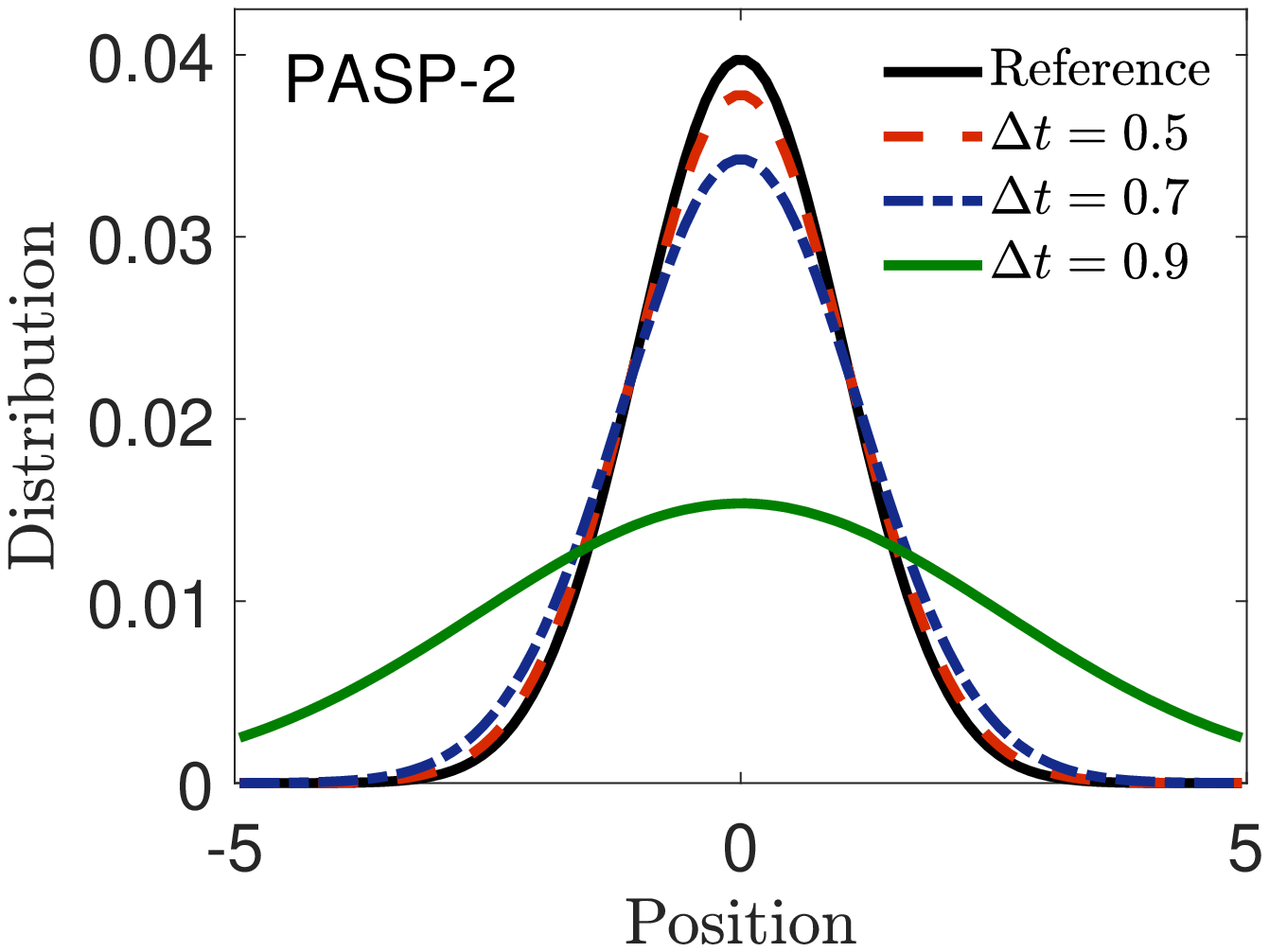}
\includegraphics[scale=0.45]{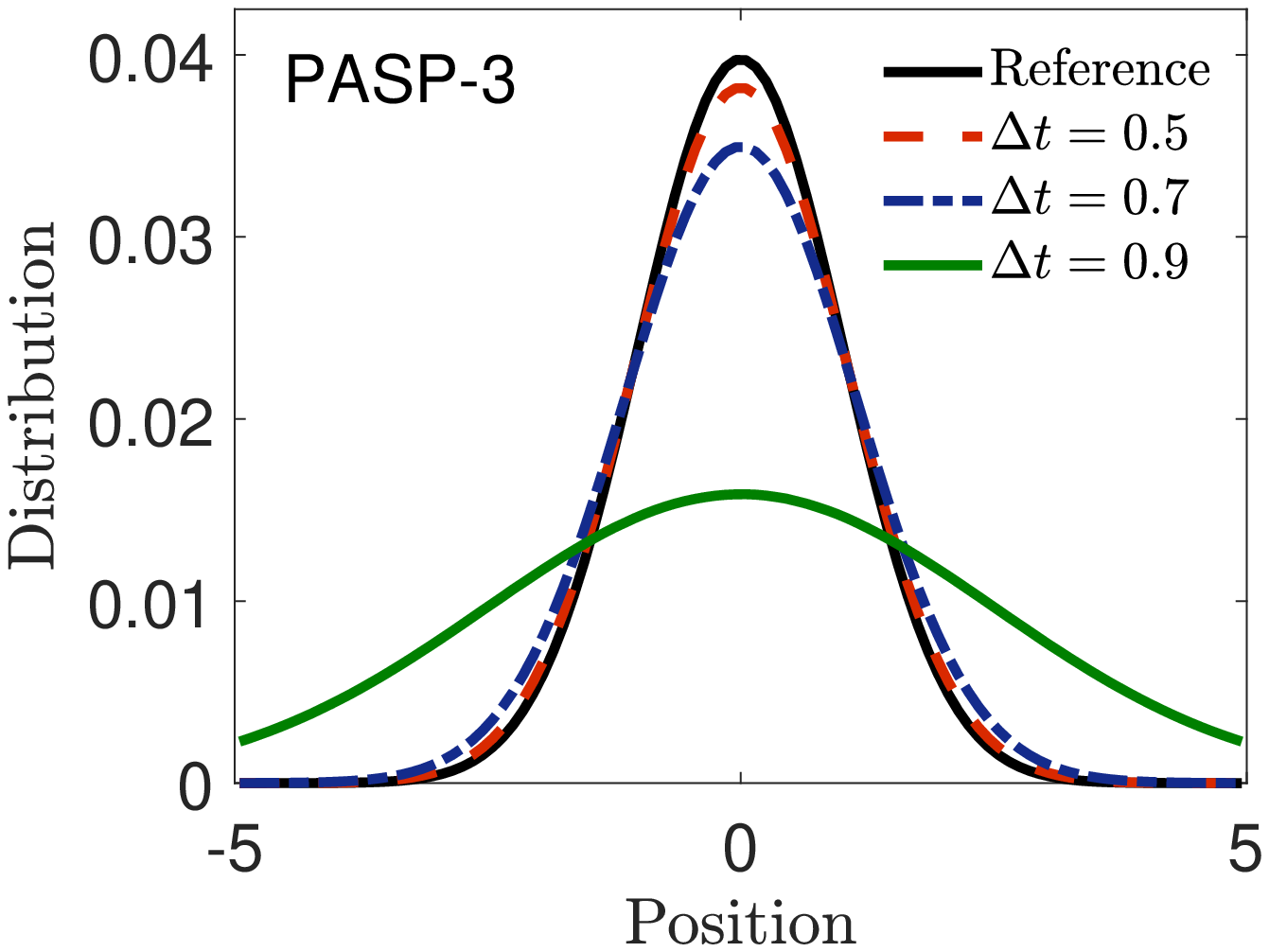}
\includegraphics[scale=0.45]{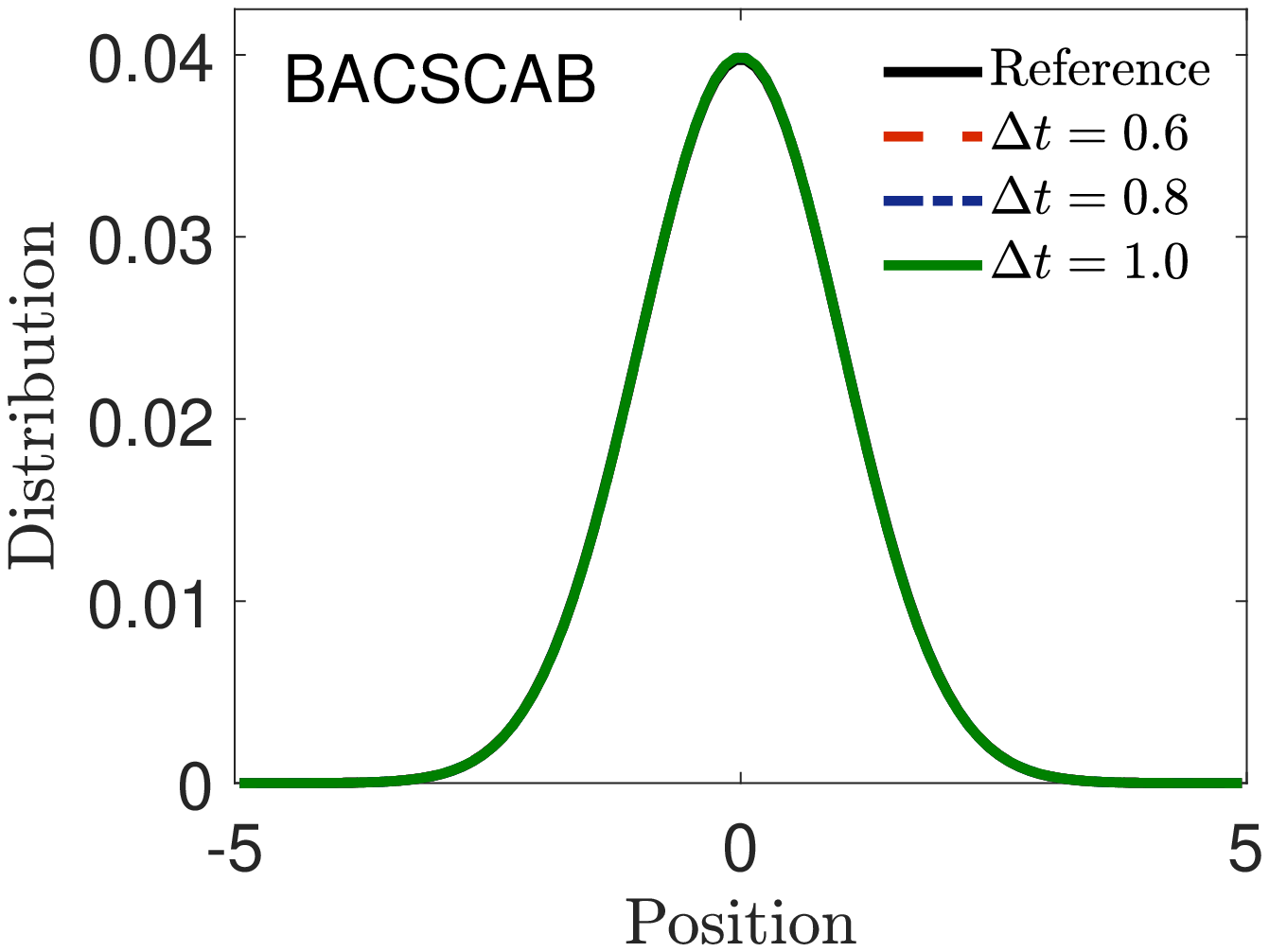}
\includegraphics[scale=0.45]{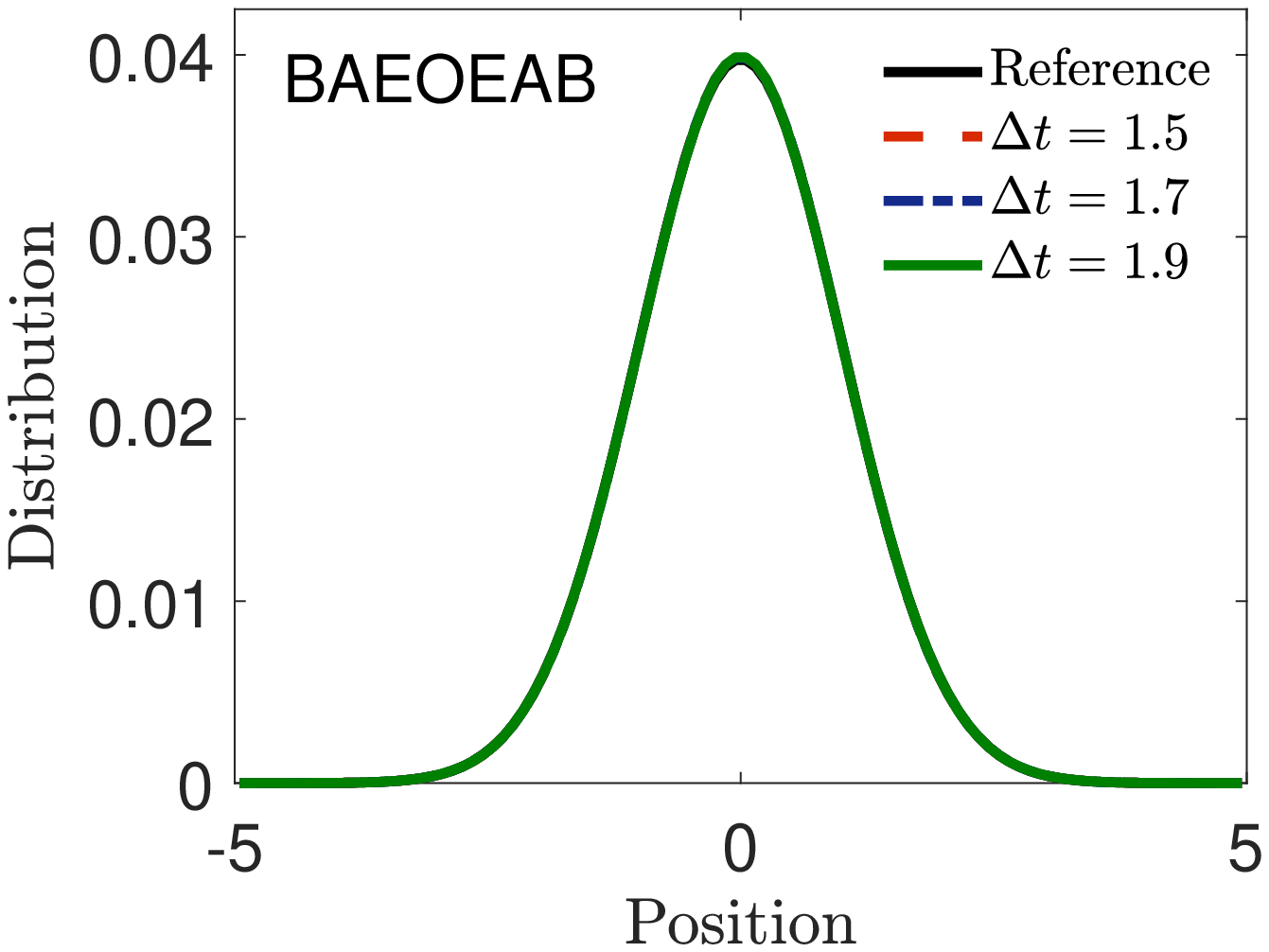}
\caption{\small Comparisons of the distributions obtained from various methods for the GLE with parameters of $\tilde{\lambda}=2$ and $\tilde{\alpha}=1$. The solid black line is the reference solution obtained by using the BACSCAB method with a very small stepsize of $\Delta t=0.01$, while the colored lines correspond to different stepsizes as indicated. (For interpretation of the colors in the figure(s), the reader is referred to the web version of this article.)}
\label{fig:GLE_HO_Comp_Dist}
\end{figure}

The distributions of positions associated with different stepsizes were compared in Figure~\ref{fig:GLE_HO_Comp_Dist}. The behavior is consistent with our findings on the left panel of Figure~\ref{fig:GLE_HO_Comp}. That is, the distributions of both PASP-2 and PASP-3 methods visibly deviate from the reference solutions with a stepsize of $\Delta t = 0.7$ and substantial deviations can be observed with a stepsize of $\Delta t = 0.9$. In contrast, the distributions of both BACSCAB and BAEOEAB methods are both indistinguishable from the reference solutions; however the largest stepsizes used in the BACSCAB and BAEOEAB methods were $\Delta t = 1$ and $\Delta t = 1.9$, respectively. This again demonstrates the superior accuracy and robustness of the BAEOEAB method over alternative methods.

\subsection{Multi-particle system}

\begin{figure}[tb]
\centering
\includegraphics[scale=0.45]{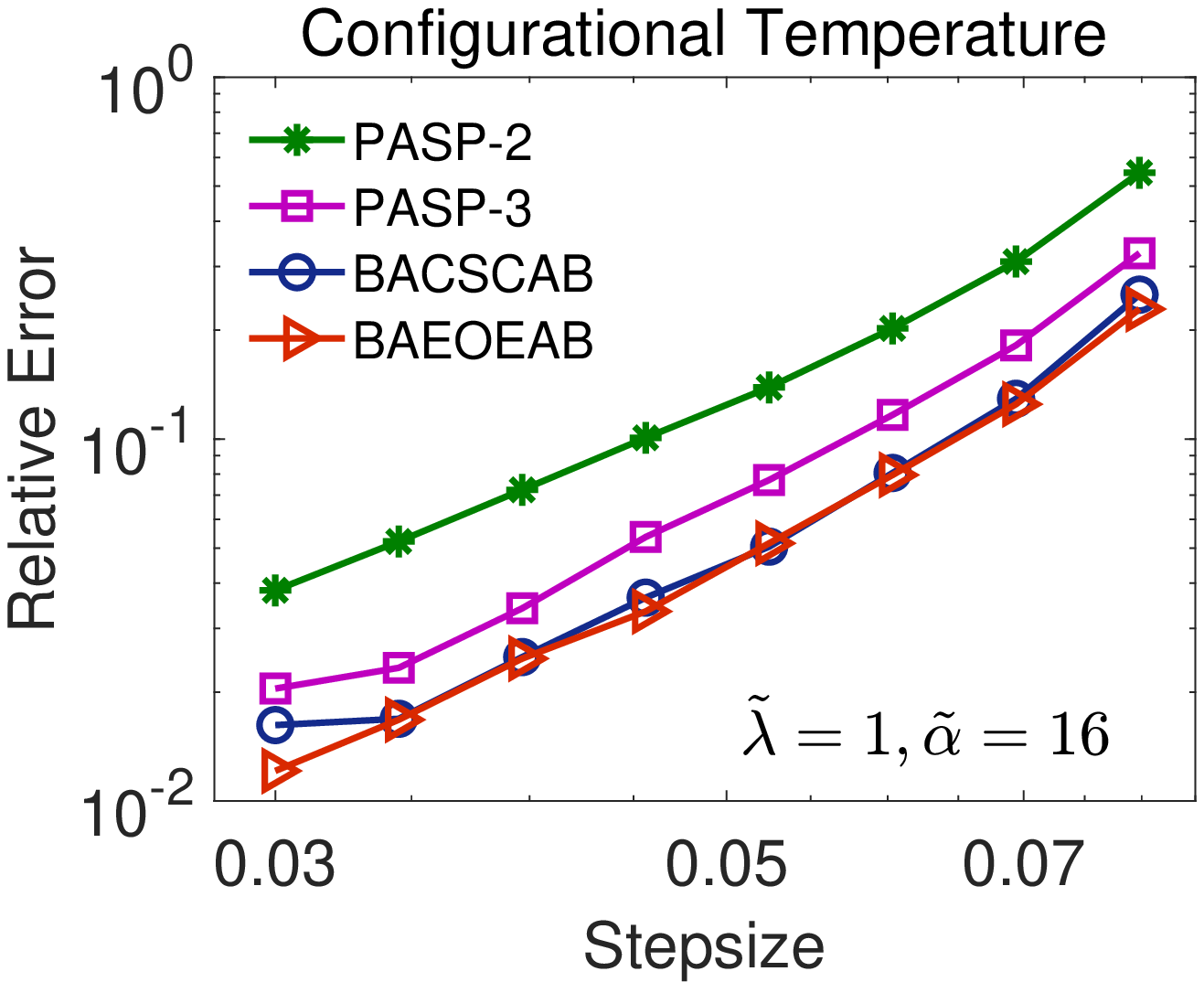}
\includegraphics[scale=0.45]{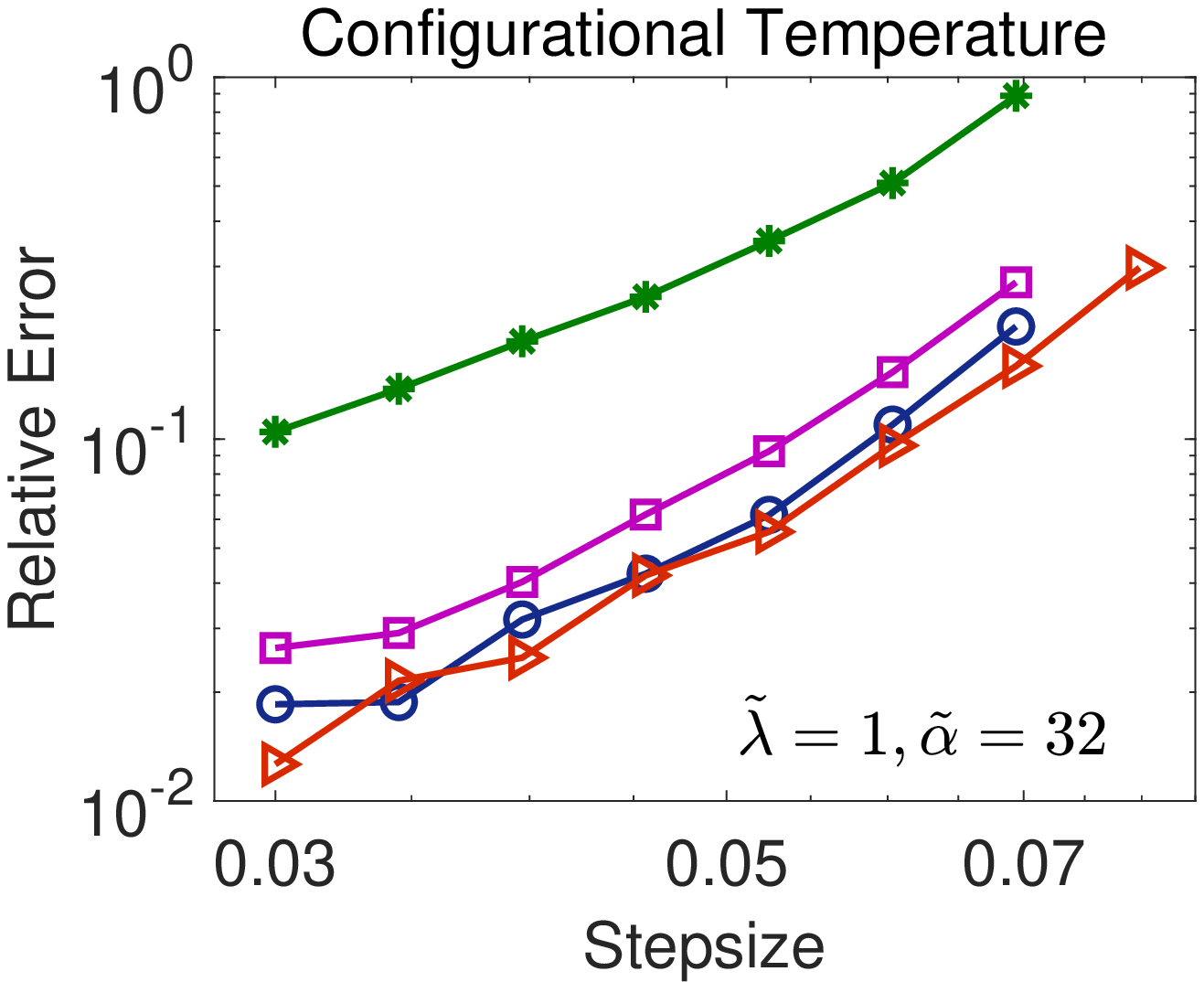}
\includegraphics[scale=0.45]{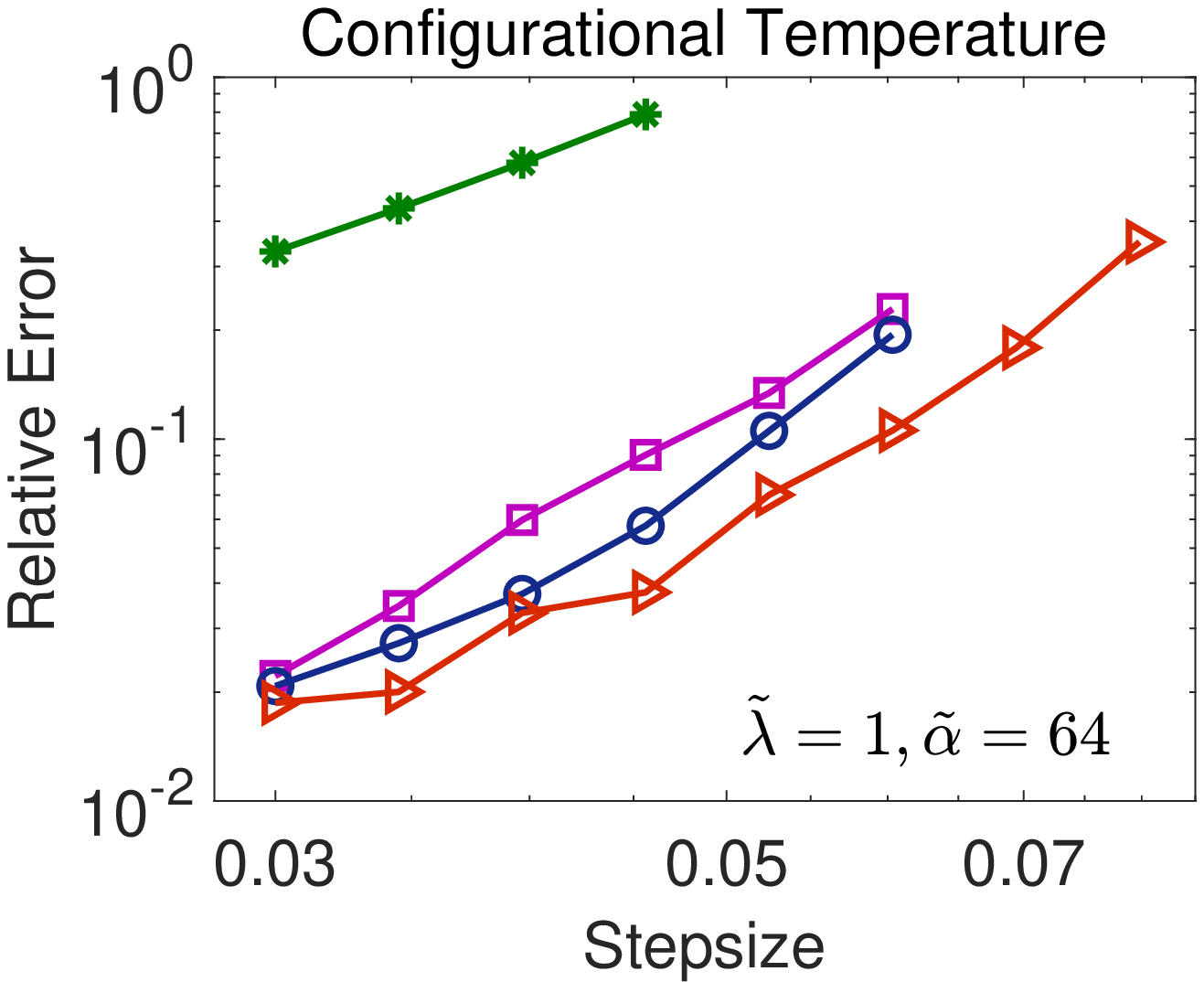}
\includegraphics[scale=0.45]{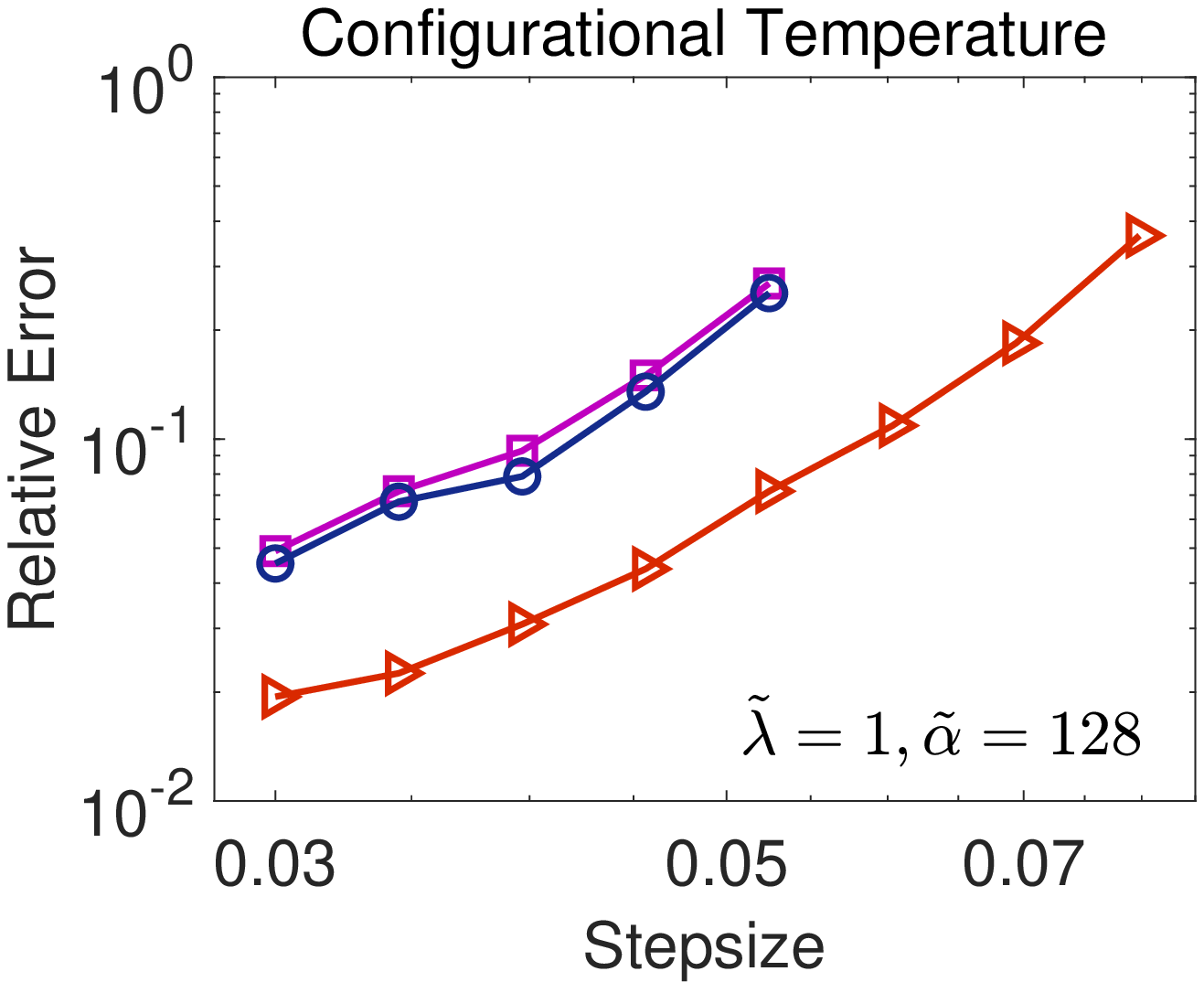}
\caption{\small Double logarithmic plot of the relative error in the computed configurational temperature against stepsize by using various numerical methods for the GLE described in Section~\ref{sec:Numerical_Methods} with parameters of $\tilde{\lambda}=1$ and $\tilde{\alpha}=16$ (top left), $\tilde{\alpha}=32$ (top right), $\tilde{\alpha}=64$ (bottom left), and $\tilde{\alpha}=128$ (bottom right). The system was simulated for $1000$ reduced time units but only the last 80\% of the data were collected to calculate the static quantity in order to make sure the system was well equilibrated. Ten different runs were averaged to reduce the sampling errors. The stepsizes tested began at $\Delta t=0.03$ and were increased incrementally by 15\% until all methods either started to show significant relative errors or became unstable. 
}
\label{fig:GLE_DPD_CT_lambda1}
\end{figure}


\begin{figure}[tb]
\centering
\includegraphics[scale=0.45]{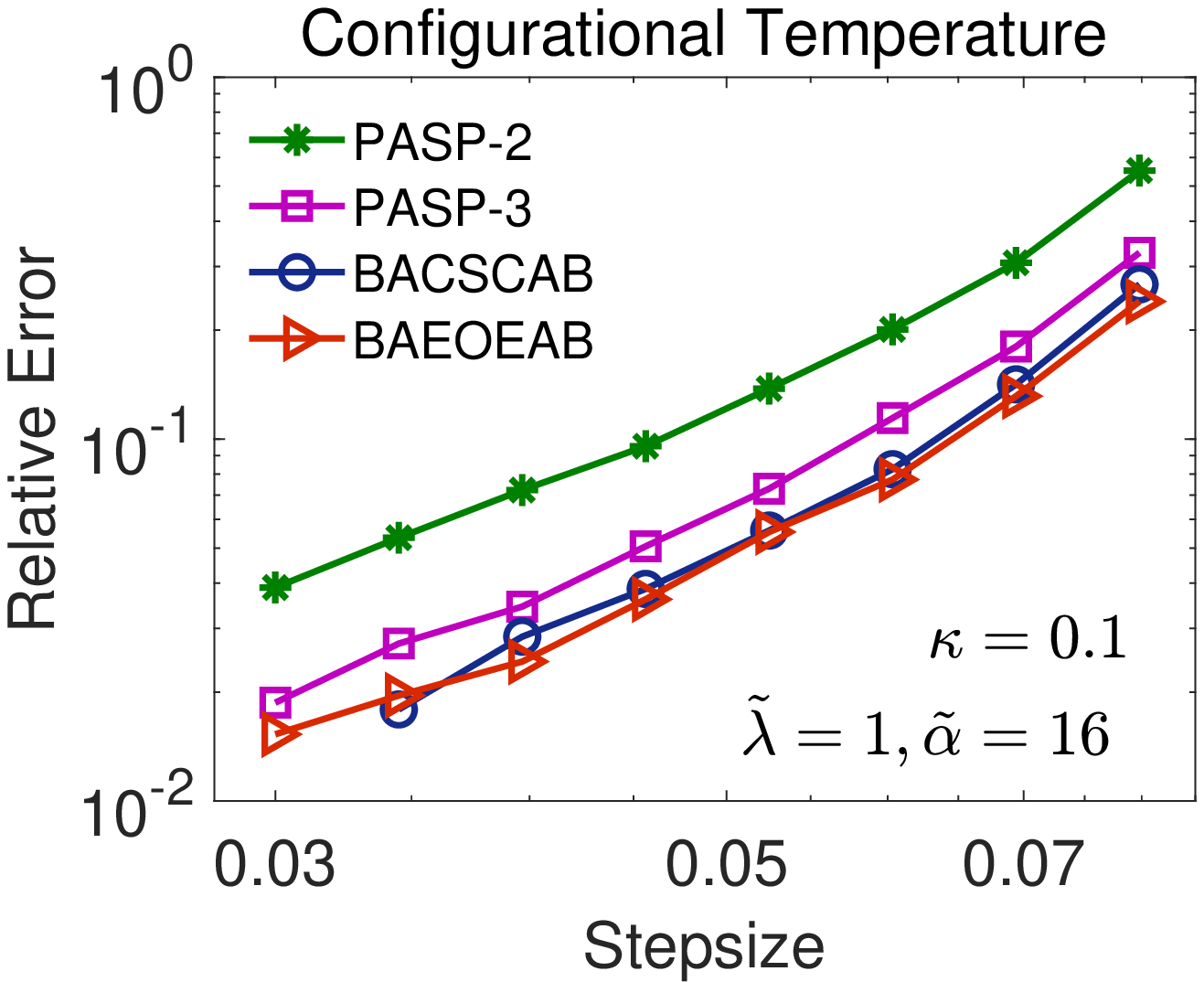}
\includegraphics[scale=0.45]{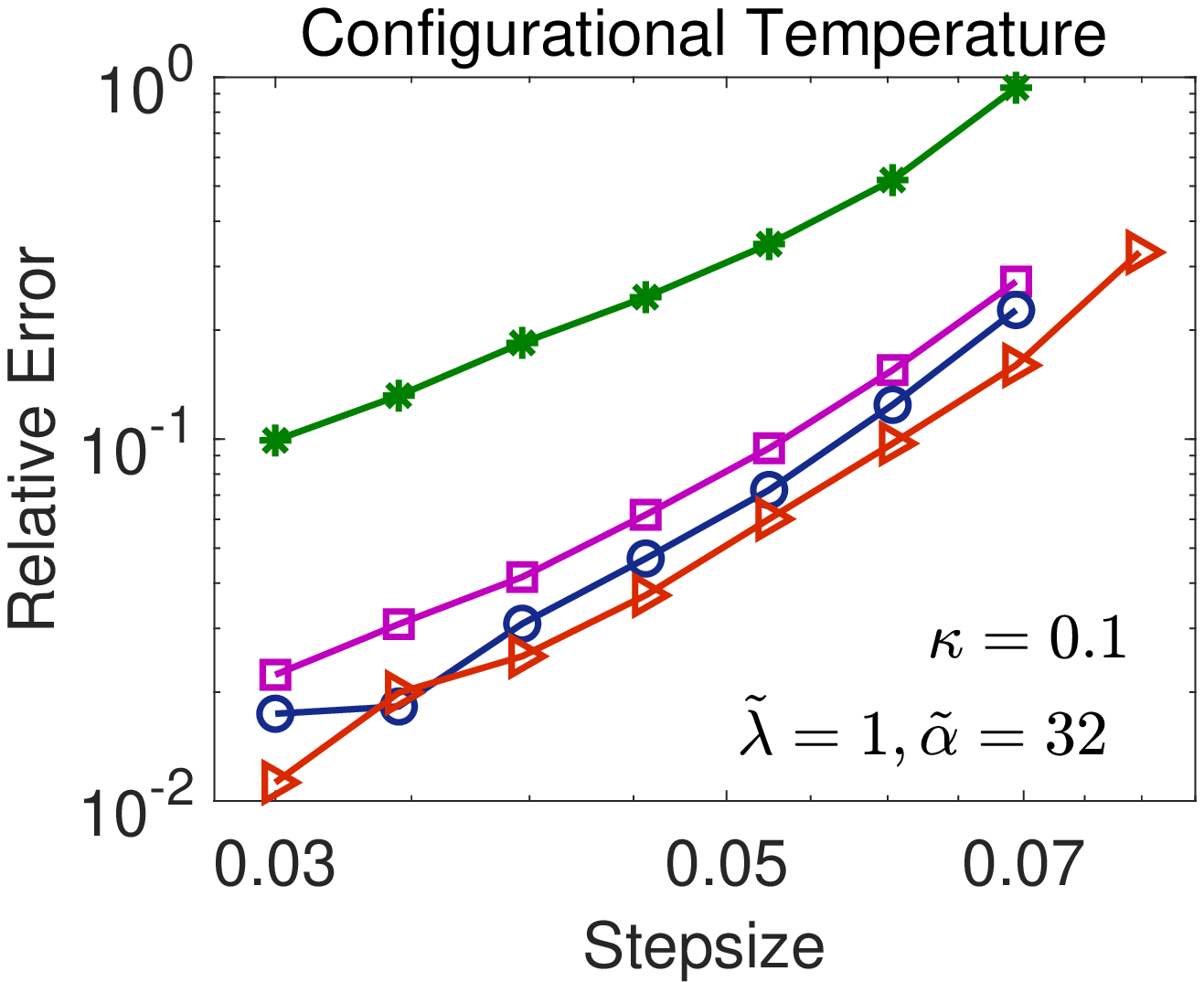}
\includegraphics[scale=0.45]{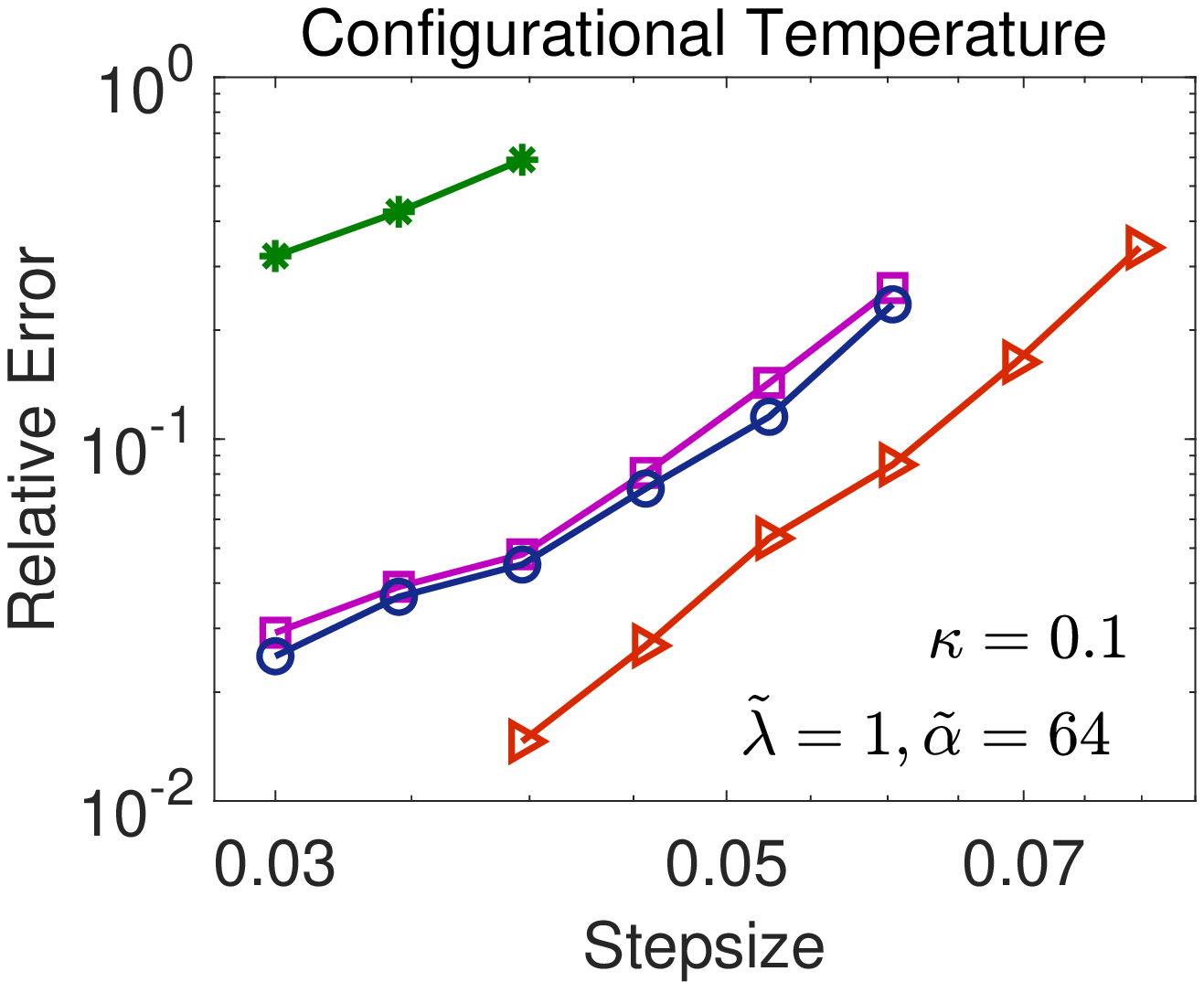}
\includegraphics[scale=0.45]{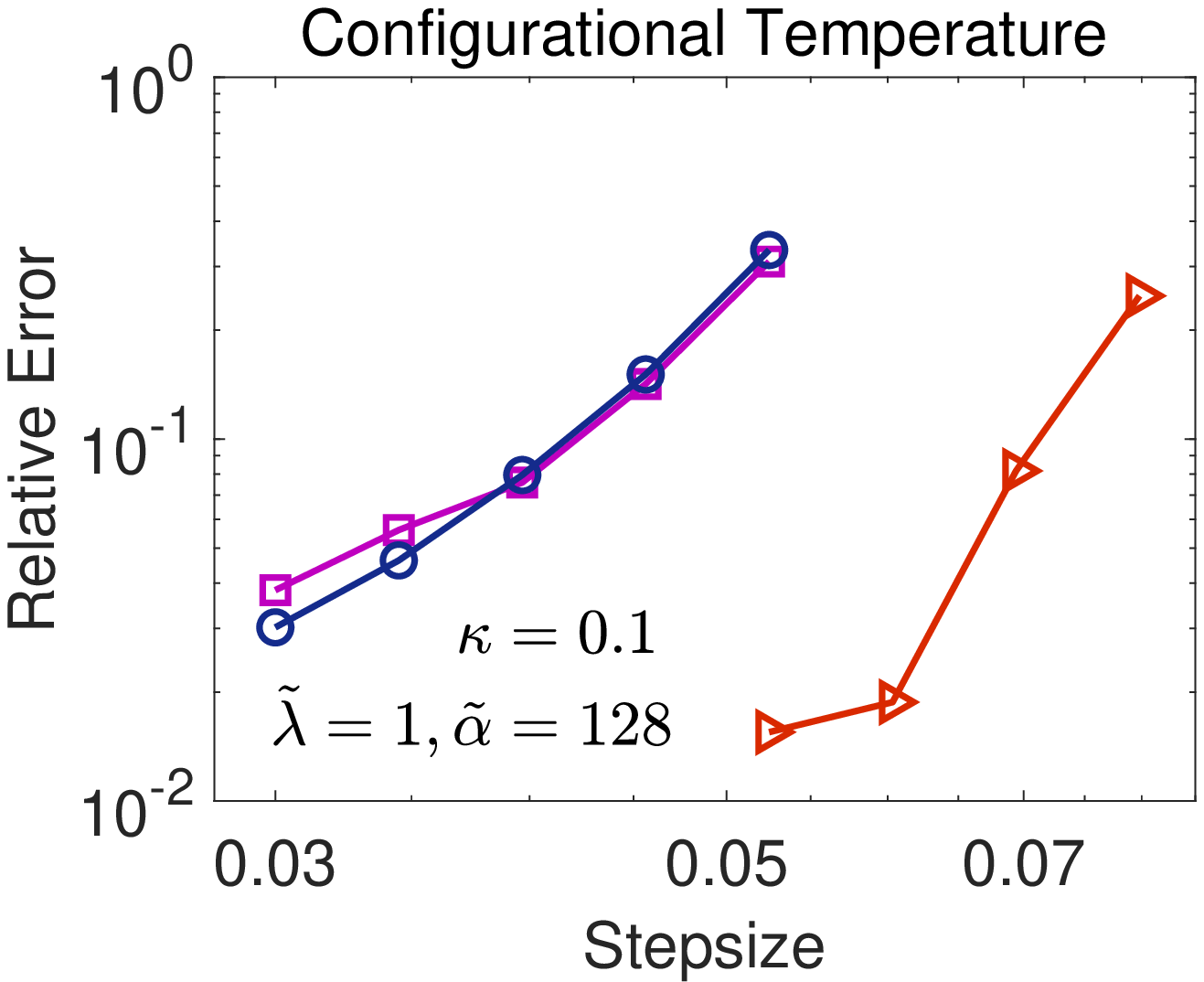}
\caption{\small Double logarithmic plot of the relative error in the computed configurational temperature against stepsize by using various numerical methods for the GLE described in Section~\ref{sec:Numerical_Methods} with parameters of $\tilde{\lambda}=1$ and $\tilde{\alpha}=16$ (top left), $\tilde{\alpha}=32$ (top right), $\tilde{\alpha}=64$ (bottom left), and $\tilde{\alpha}=128$ (bottom right). The format of the plots is the same as in Figure~\ref{fig:GLE_DPD_CT_lambda1}, except with a shear rate of $\kappa=0.1$ generated by the Lees--Edwards boundary conditions.
}
\label{fig:GLE_DPD_CT_lambda1_kappad1}
\end{figure}

We also compare the performance of various methods in systems where multiple particles are interacting with each other. To this end, we chose a soft pair potential energy that has been widely used in the so-called dissipative particle dynamics~\cite{Leimkuhler2015,Leimkuhler2016a,Shang2020},
\begin{equation}
  U(\mathbf{q}) = \sum_{i} \sum_{j>i}\varphi(r_{ij}) \, ,
\end{equation}
where
\begin{equation}\label{eq:Pair_Potential}
  \varphi(r_{ij})=
  \begin{cases}
    a_{ij}r_{\mathrm{c}} \left(1-r_{ij}/r_{\mathrm{c}}\right)^{2}/2 \, , & r_{ij} < r_{\mathrm{c}} \, ,\\
    \quad \quad \quad \quad 0 \, , & r_{ij} \geq r_{\mathrm{c}} \, ,
  \end{cases}
\end{equation}
where parameter $a_{ij}$ denotes the maximum repulsion strength between particles $i$ and $j$, $r_{ij} = |\mathbf{q}_{ij}| = |\mathbf{q}_{i}-\mathbf{q}_{j}|$ is the distance, and $r_{\mathrm{c}}$ represents a certain cutoff radius, beyond which there is no interaction between particles. We adopted a standard set of parameters commonly used in algorithms tests as in~\cite{Shang2020}: $m_{i} = r_{\mathrm{c}} = \beta = 1$. Moreover, a particle density of $\rho_{\rm d}=3$ was used throughout the current article, which also determines the repulsion parameter of $a_{ij} = 75/(\beta\rho_{\rm d}) = 25$ in order to match the compressibility of water~\cite{Groot1997}. A system of $N=500$ identical particles was simulated in a cubic box with periodic boundary conditions~\cite{Allen2017}, unless otherwise stated. The initial positions of the particles were i.i.d.\ with a uniform distribution over the box, while the initial momenta were i.i.d.\ normal random variables with mean zero and variance $1/\beta$.

The average of the computed configurational temperature in the canonical ensemble is expected to be precisely the target temperature:
\begin{equation}\label{eq:Config_Temp}
  \beta^{-1}
  = \frac{  \left\langle \nabla_{i} U(\mathbf{q}) \cdot \nabla_{i} U(\mathbf{q}) \right\rangle }{ \left\langle \nabla^{2}_{i}U(\mathbf{q}) \right\rangle } \, ,
\end{equation}
where $\nabla_{i}U$ and $\nabla^{2}_{i}U$, respectively, represent the gradient and Laplacian of the potential energy $U$ with respect to the position of particle $i$. The control of the configurational temperature has been recommended in~\cite{Allen2006} as a verification of equilibrium; importantly, good control of the configurational temperature also appears to imply good performance in other configuration-based physical quantities (see more discussions on the configurational temperature in~\cite{Leimkuhler2015,Leimkuhler2016a,Shang2020}).

The control of the configurational temperature was compared in Figure~\ref{fig:GLE_DPD_CT_lambda1} with $\tilde{\lambda}=1$ and varying $\tilde{\alpha}$. All the methods appear to have second order convergence to the invariant measure (dashed order lines not shown). On the top left panel where $\tilde{\alpha}=16$, the PASP-3 method appears to be more accurate than the PASP-2 method; however it is outperformed by either BACSCAB or BAEOEAB methods, whose curves are almost indistinguishable. As we increase the value of $\tilde{\alpha}$, the behavior is largely similar except (a) PASP-2 becomes less and less accurate than alternative methods and displays substantial relative error at a stepsize of around $\Delta t = 0.05$ on the bottom left panel where $\tilde{\alpha}=64$; (b) BAEOEAB becomes increasingly better than alternative methods in terms of not only accuracy but also robustness---while, on the bottom right panel where $\tilde{\alpha}=128$, both PASP-3 and BACSCAB show an around 25\% relative error with a stepsize of slightly over $\Delta t = 0.05$, the relative error is around 36\% for BAEOEAB with a stepsize of around $\Delta t = 0.08$.

Figure~\ref{fig:GLE_DPD_CT_lambda1_kappad1} also compares the control of the configurational temperature when the well-known Lees--Edwards boundary conditions~\cite{Lees1972} were applied in order to generate a simple and steady shear flow (shear rate $\kappa=0.1$), again with $\tilde{\lambda}=1$ and varying $\tilde{\alpha}$. The behavior is very similar to Figure~\ref{fig:GLE_DPD_CT_lambda1} except the superiority of the BAEOEAB method is even more evident particularly in the bottom row where $\tilde{\alpha}$ is relatively large. For instance, on the bottom right panel where $\tilde{\alpha}=128$, BAEOEAB achieves around an order of magnitude improvement over both \mbox{PASP-3} and BACSCAB in terms of accuracy. This indicates that BAEOEAB has a even better configurational temperature control over alternative methods in nonequilibrium simulations than in equilibrium when $\tilde{\alpha}$ is relatively large. It is worth mentioning that, although BAEOEAB has been designed in equilibrium settings where the invariant measure is preserved, it appears that its performance is also very good in nonequilibrium, particularly when the shear rate is relatively small, which may be thought of as in ``near-equilibrium''.

\section{Conclusions}
\label{sec:Conclusions}

We have reviewed the constructions of popular splitting methods proposed for the GLE. Having also pointed the potential drawback of those existing methods, we have proposed an alternative method based on a different splitting of the vector field of the GLE. We have also analyzed the errors on the averages associated with different methods in the case of a one-dimensional harmonic oscillator. We have demonstrated that all of the averages are exact for the newly proposed BAEOEAB method, with the only exception being the average of $\langle p^{2} \rangle$ that has second order errors, leading to a friction-independent upper bound of the stepsize $\Delta t_{\mathrm{max}}=2\sqrt{m/K}$ that coincides with the (deterministic) Verlet stability threshold. In contrast, there are at least two averages that are not exact for all the other splitting methods examined.

Restricting our attention to a single mode (i.e., $M=1$) for simplicity, we have systematically compared the BAEOEAB method with alternative methods in a variety of numerical experiments. In the case of a one-dimensional harmonic oscillator, the BAEOEAB method is exact in both averages of $\langle q^{2} \rangle$ and $\langle z^{2} \rangle$ and substantially outperforms alternative methods in terms of not only accuracy but also robustness. The BACSCAB method is also exact in the average of $\langle q^{2} \rangle$, and appears to be as accurate as the BAEOEAB method. However, the former became unstable just over $\Delta t = 1$, while the latter is still extremely accurate (i.e., up to sampling error) when the stepsize is around $\Delta t = 1.9$, close to its stability threshold $\Delta t_{\mathrm{max}}=2$. In the average of $\langle z^{2} \rangle$, the BACSCAB method becomes second order as both PASP methods. Moreover, the performance of the three methods is very similar to each other, all outperformed by the BAEOEAB method.

We have also examined the case of multiple particles that are interacting with each other---a soft pair potential energy widely used in the so-called dissipative particle dynamics. Fixing $\tilde{\lambda}$ to be unity, as we increase the value of $\tilde{\alpha}$, the BAEOEAB method becomes increasingly better than alternative methods in terms of not only accuracy but also robustness. Moreover, the superiority of the BAEOEAB method is even more evident when the well-known Lees--Edwards boundary conditions were applied in order to generate a simple and steady shear flow, a technique commonly used in nonequilibrium simulations.

Although we have focused our attention on a single mode in this article, it is expected that the newly proposed BAEOEAB method will perform well in multi-mode cases (i.e., a general Prony series) that can be viewed as a summation of single modes. However, we leave a detailed examination (including more general forms of the memory kernel, for instance, in Section~\ref{subsec:Markovian_Approximations}) for future work.

Following similar results in Langevin dynamics~\cite{Leimkuhler2013,Leimkuhler2015a}, a fourth-order convergence to the invariant measure has been demonstrated recently for a particular splitting method for the GLE with certain choices of the parameters~\cite{Leimkuhler2020}. However, such superconvergence results have not been generally observed in our numerical experiments. Therefore, we leave further exploration of this observation for future work.

\section*{CRediT authorship contribution statement}

\textbf{Manh Hong Duong:} Conceptualization, Investigation, Methodology, Writing - review \& editing. \textbf{Xiaocheng Shang:} Conceptualization, Formal analysis, Investigation, Methodology, Software, Validation, Visualization, Writing - original draft, Writing - review \& editing.

\section*{Declaration of competing interest}

The authors declare that they have no known competing financial interests or personal relationships that could have appeared to influence the work reported in this paper.

\section*{Acknowledgements}

The authors thank Nawaf Bou-Rabee, Benedict Leimkuhler, Matthias Sachs, and the anonymous referees for their valuable suggestions and comments. XS acknowledges the support of the London Mathematical Society through the Research Reboot Grant (reference number 42025), the Royal Society through the International Exchanges Scheme (reference number IES$\backslash$R3$\backslash$203007), and the Institute of Mathematics and its Applications through the QJMAM Fund for Applied Mathematics. The research of MHD was supported by EPSRC Grants EP/W008041/1 and EP/V038516/1.


\appendix






\bibliographystyle{is-abbrv}

\bibliography{refs}

\end{document}